\documentclass{article}

\usepackage{PRIMEarxiv}
\usepackage{comment}
\usepackage[utf8]{inputenc} 
\usepackage[T1]{fontenc}    
\usepackage{hyperref}       
\usepackage{url}            
\usepackage{booktabs}       
\usepackage{amsfonts}       
\usepackage{nicefrac}       
\usepackage{microtype}      
\usepackage{lipsum}
\usepackage{fancyhdr}       
\usepackage{graphicx}       
\graphicspath{{media/}}     

\newcommand{\wse}[0]{WSE}
\newcommand{\ProcElem}[0]{PE}
\newcommand{\NetOnChp}[0]{NoC}
\newcommand{\ExecPE}[0]{E-PE}
\newcommand{\RespPE}[0]{R-PE}
\newcommand{\MergePE}[0]{M-PE}
\newcommand{\AbsSynTree}[0]{AST}
\newcommand{\RemProc}[0]{RPC}
\newcommand{\DomSpecLang}[0]{DSL}
\newcommand{\GSclr}[0]{GS}
\newcommand{\GArr}[0]{GA}
\newcommand{\LSclr}[0]{LS}
\newcommand{\LArr}[0]{LA}
\newcommand{\ULSclr}[0]{ULS}
\newcommand{\IntLang}[0]{IL}
\newcommand{\IntRepGraph}[0]{IRG}
\newcommand{\graphInterp}[0]{IRG-I}
\newcommand{\oods}[0]{OODS}
\newcommand{\virtMachine}[0]{VM}
\newcommand{\contextMan}[0]{CM}
\newcommand{\mpi}[0]{MP}

\newcommand{\PaperTitle}[0]{A System Level Compiler for Massively-Parallel, Spatial, Dataflow Architectures}

\usepackage{xcolor, colortbl}

\definecolor{codegreen}{rgb}{0,0.6,0}
\definecolor{codegray}{rgb}{0.5,0.5,0.5}
\definecolor{codepurple}{rgb}{0.58,0,0.82}
\definecolor{backcolour}{rgb}{0.95,0.95,0.92}
\definecolor{main-color}{rgb}{0.6627, 0.7176, 0.7764}
\definecolor{back-color}{rgb}{0.1686, 0.1686, 0.1686}
\definecolor{string-color}{rgb}{0.3333, 0.5254, 0.345}
\definecolor{key-color}{rgb}{0.8, 0.47, 0.196}
\definecolor{table_row}{rgb}{0.9,0.9,0.9}

\usepackage{listings}
\usepackage{float}
\newfloat{lstfloat}{htbp}{lop}
\floatname{lstfloat}{Listing}

\lstdefinestyle{pythonstyle}{
    backgroundcolor=\color{backcolour},   
    commentstyle=\color{codegreen},
    keywordstyle=\color{magenta},
    numberstyle=\tiny\color{codegray},
    stringstyle=\color{codepurple},
    basicstyle=\ttfamily\footnotesize,
    breakatwhitespace=false,         
    breaklines=true,                 
    captionpos=b,                    
    keepspaces=true,                 
    numbers=left,                    
    numbersep=5pt,                  
    showspaces=false,                
    showstringspaces=false,
    showtabs=false,                  
    tabsize=2,
    xleftmargin= 0.15in
}

\lstdefinestyle{cppstyle}
{
    language = C++,
    backgroundcolor=\color{backcolour},
    commentstyle=\color{codegreen},
    numberstyle=\tiny\color{codegray},
    basicstyle = \ttfamily\footnotesize,  
    stringstyle = {\color{string-color}},
    keywordstyle = {\color{key-color}},
    keywordstyle = [2]{\color{key-color}},
    keywordstyle = [3]{\color{yellow}},
    keywordstyle = [4]{\color{teal}},
    morekeywords = [2]{forever,in,dispatch,socket,param,parallel},
    morekeywords = [3]{<<, >>},
    morekeywords = [4]{sp,xp,wv},
    breakatwhitespace=false,         
    breaklines=true,                 
    captionpos=b,                    
    keepspaces=true,                 
    numbers=left,                    
    numbersep=5pt,                  
    showspaces=false,                
    showstringspaces=false,
    showtabs=false,                  
    tabsize=2,
    xleftmargin= 0.15in
}

\lstdefinestyle{paint}
{
    language = C++,
    backgroundcolor=\color{backcolour},
    commentstyle=\color{codegreen},
    numberstyle=\tiny\color{codegray},
    string = [b]{`}
    basicstyle = \ttfamily\footnotesize,  
    keywordstyle = {\color{key-color}},
    keywordstyle = [2]{\color{key-color}},
    keywordstyle = [3]{\color{yellow}},
    keywordstyle = [4]{\color{teal}},
    morekeywords = [2]{let,color, define, tile,vstack,hstack,hstackrep,vstackrep,paint,split,ring,code},
    morekeywords = [3]{<<, >>},
    morekeywords = [4]{sp,xp,wv},
    breakatwhitespace=false,         
    breaklines=true,                 
    captionpos=b,                    
    keepspaces=true,                 
    numbers=left,                    
    numbersep=5pt,                  
    showspaces=false,                
    showstringspaces=false,
    showtabs=false,                  
    tabsize=2,
    xleftmargin= 0.15in
}

\usepackage{makecell}
\usepackage{tabularray}
\UseTblrLibrary{booktabs}
\usepackage[font=small,labelfont=bf]{caption}

\title{\PaperTitle
\thanks{\textit{\underline{Citation}}: 
\textbf{D. Van Essendelft, P. Wingo, T. Jordan, R. Smith,  and W. Saidi \PaperTitle. 26 Pages DOI:TBD}} 
}

\author{
  Dirk Van Essendelft$^{\diamondsuit\dag}$, Patrick Wingo$^\ddag$, Terry Jordan$^\dag$, Wissam A. Saidi$^\P$\\
  U.S. Department Of Energy \\
  The National Energy Technology Laboratory \\
  $^\dag$ Morgantown, WV, USA\\
  $^\ddag$ Albany, OR, USA\\
  $^\P$ Pittsburgh, Pa, USA\\
  \texttt{$^{\diamondsuit}$dirk.vanessendelft@netl.doe.gov} \\
   \And
  Ryan Smith \\
  U.S. Department Of Energy \\
 The National Energy Technology Laboratory \textendash\ Postdoctoral Research Fellowship Program \\
  Pittsburgh, PA, USA\\
}

\begin{document}
\maketitle

\begin{abstract}
We have developed a novel compiler called the Multiple-Architecture Compiler for Advanced Computing Hardware (MACH) designed specifically for massively-parallel, spatial, dataflow architectures like the Wafer Scale Engine. Additionally, MACH can execute code on traditional unified-memory devices. MACH addresses the complexities in compiling for spatial architectures through a conceptual Virtual Machine, a flexible domain-specific language, and a compiler that can lower high-level languages to machine-specific code in compliance with the Virtual Machine concept. While MACH is designed to be operable on several architectures and provide the flexibility for several standard and user-defined data mappings, we introduce the concept with dense tensor examples from NumPy and show lowering to the Wafer Scale Engine by targeting Cerebras' hardware specific languages.
\end{abstract}


\newpage

\section{Introduction}\label{introduction}

\begin{figure}[h!]
    \centering
  \includegraphics[width=\textwidth]{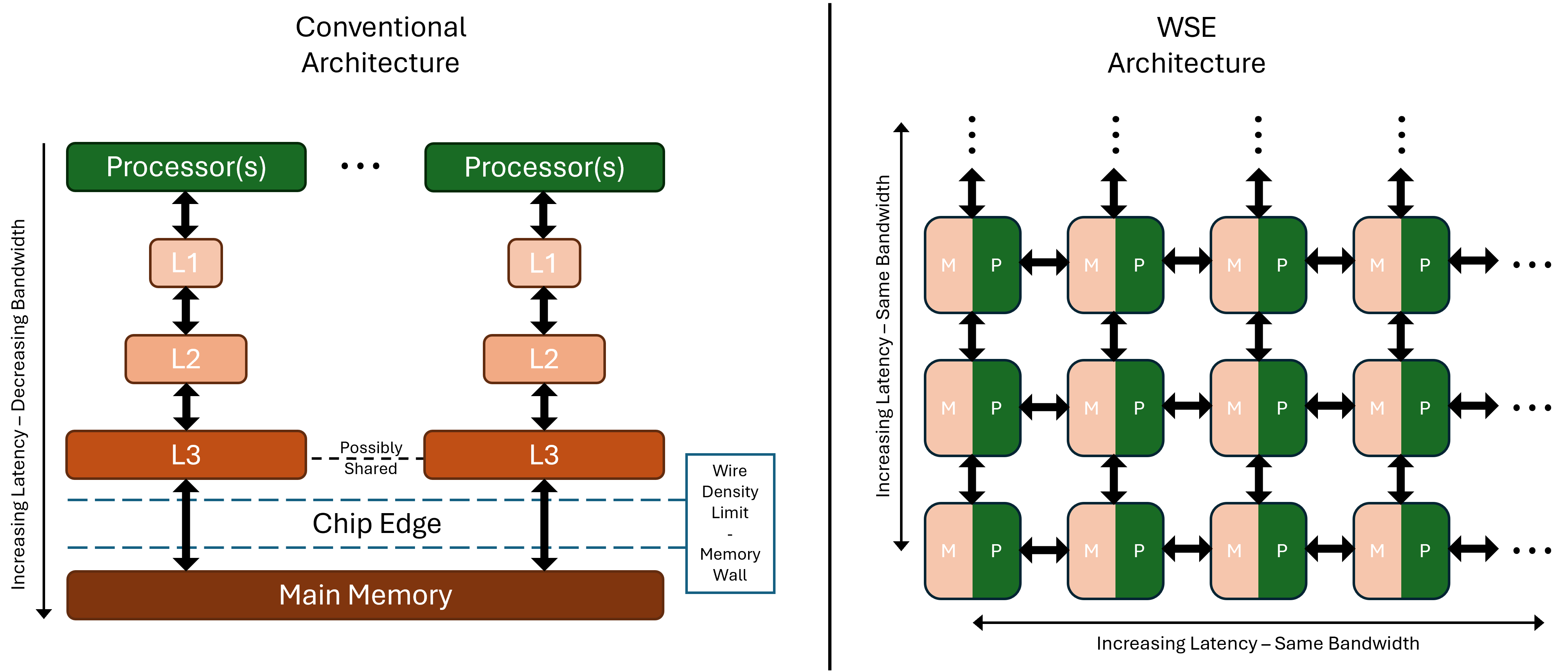}
    \caption{Conventional vs \wse\  Architecture. In Conventional architectures (left), processors have to reach vertically through a memory pool that is substantially larger than L1 cache to access the main memory shared between processors. The last layer of cache may be shared between some or all processors (i.e on a GPU) before having to reach off the processor edge to main memory. Latency increases with depth in the hierarchy as bandwidth drops. There is a substantial increase in latency and decrease in bandwidth to reach off the chip edge. L1 cache is usually the only memory close enough to the processor to supply 2 or 3 operands on read and 1 on write at every cycle, matching it to process rate. The \wse\ (right) has no cache hierarchy. Each PE has a processor and memory cache, which eliminates memory wall within the \NetOnChp\ reach. \ProcElem s are arranged in a rectilinear Cartesian layout. Communication bandwidth (bi-directional arrows) between \ProcElem s on the \NetOnChp\ is matched to process rate such that one operand can be sent/sourced to/from the \NetOnChp. One or two cycles of latency are incurred for every hop on the \NetOnChp\ fabric, but bandwidth is maintained. Messaging farther than to nearest neighbors proceeds by hopping across the \NetOnChp\ fabric. The latency is relative to the starting and ending point of the message sent. }
    \label{fig:VNVsSpatial}
\end{figure}

The escalating computational demands of artificial intelligence (AI), have prompted the development of novel computing architectures that significantly depart from traditional processor-memory designs \cite{Lie_HotChips, Dojo_micro, Sambanova_micro, Grok_Micro, tenstorrent_docs}. Conventional architectures have one or more processors that can access a large memory pool through several layers of cache (Fig. \ref{fig:VNVsSpatial}, left). Massively-parallel, spatial, data-flow architectures, like the Wafer Scale Engine (\wse), offer high throughput by distributing processing and memory across a large array of interconnected Processing Elements (\ProcElem s). Each \ProcElem\ on the \wse\ is composed of a processor, memory, and a Network-on-Chip (\NetOnChp) router that allows communication to adjacent \ProcElem s (Fig. \ref{fig:VNVsSpatial}, right) \cite{Lie_HotChips}. Importantly, there is no wire density limit or long physical data path between a processor and its memory which enables the memory system to keep up with the processing rate provided the data and its processor are roughly co-located. In this way, spatial architectures trade a single, large, relatively slow memory pool for a large number of small but very fast memory pools located on each \ProcElem. This design establishes a very fast memory system that is unified at the \ProcElem\ level but not at the system level. As a result, a \emph{near-memory} computing environment exists that can mitigate the memory wall when data and a processor are nearby within the spatial architecture. In this environment, bandwidth is maintained at processor consumption rates across the system but latency increases (1-2 cycles) with each \ProcElem\ hop on the \NetOnChp. It is often the case that scientific models can be significantly accelerated within this computing environment because a decent mapping that preserves data locality can often be found \cite{disruptiveChanges, rocki2020fast, sai2024matrix, sai2023massively, recordIsing, mdWSE, mcParticleTransport}. Good mappings for a variety of scientific models likely exist because of the symmetry between principle of locality in physics \cite{Haag1992} and the processor-memory locality characteristics of spatial architectures. Due to the principle of locality in physics, the time evolution of complex global behavior is determined by a series of local events (particle collisions and oscillations of fields for instance). This often means that models can be partitioned in a suitable way to take advantage of rapid memory access within a spatial architecture.

However, the non-unified memory structure inherent in spatial architectures presents a significant compiler challenge. Unlike conventional systems where simple flattening techniques allow general data placement and access by any processor connected to the memory pool, spatial architectures require explicit data placement across the grid of \ProcElem s. This requires careful management of the limited memory on each \ProcElem, explicit management of data movement between \ProcElem s, and a coordinated approach to instructing individual \ProcElem s to execute a program at the system level. Furthermore, processor-data locality is a critical factor for efficient utilization in spatial architectures, a concern that is significantly less pronounced in conventional architectures.

\section{Key Compiler Concepts}
To overcome these challenges, we are developing a compiler called the Multiple-Architecture Compiler for Advanced Computing Hardware (MACH) \cite{mach_program}. As shown in Table \ref{tab:innovations}, MACH uses a novel system-level approach comprising: 1) a hardware agnostic Virtual Machine (\virtMachine), 2) a physical mapping to a hardware architecture, 3) a flexible Domain Specific Language (\DomSpecLang) compatible with the \virtMachine, and 4) a compiler to transform high-level programs into a set of programs that operate in accordance with the \virtMachine\ concept. This framework was specifically designed to provide structure and operating rules that enable the \virtMachine\ to operate in a unified manner on finely distributed on spatial architectures, but can also operate in a unified-memory environment. As a first demonstration of this system, we have focused on compiling codes written in NumPy \cite{harris2020array} for the \wse. Each of these components will be discussed in detail with code examples.

\begin{table}[h!]
    \centering
    \caption{Concepts, Purpose, and Components}
    \label{tab:innovations}

    \begin{tblr}{colspec = {Q[c,m] Q[c,m] Q[c,m]},
                 row{1} = {font=\bfseries, bg=gray!50},
                 row{odd[2]} = {bg=gray!25},
                 }
        \toprule
        \SetRow{}
        Concept         &   Purpose & Components\\
        
        \midrule
        
        Virtual Machine &   Establish Roles and Responsibilities & {Controller / Worker Strategy \\ Procedural Programming System \\ Division of Data} \\

        {Physical \ProcElem\ \\ Virtual Machine \\ Mapping} & {Processors and Memory Role Assignment \\ Machine Specific Role Implementation} & {Spatial Division of \ProcElem\ Role Assignment \\ Spatially Centralized Control System \\ Multi-\ProcElem, Vectorized Control System \\ Dedicated Reduction System \\ Moat Free Operation}\\

        {Domain \\ Specific \\ Language} & {Establish Valid Data Structures \\ Relate Data Structures to Processors and Memory \\ Establish/Arbitrate Valid Operations \\ Provide High-level Programmability \\ Establish Standardized Intermediaries \\ Compile High Level Languages to Intermediaries}  & {Object Oriented Data Structures \\ High Level Language Front End Compiler \\ Intermediate Language \\ Intermediate Graph Representation \\ Memory Manager} \\

        {Machine \\ Specific \\ Compiler} & {Compile Intermediaries to Machine Language \\ Implement Physical Mapping \\ Manage Data Placement \\ Manage Communication} & {Graph to Code Translators \\
        Code Templates \\ Compiler Variables \\ Compiler Directives \\ Optimizers}\\
        
        \bottomrule
    \end{tblr} 
\end{table}

\section{The Virtual Machine}\label{sec:virtual_machine}

We start by defining a conceptual, hardware agnostic Virtual Machine (\virtMachine). The \virtMachine\ is composed of a control system and one or more workers, as shown in Table \ref{tab:VM_Roles}. The control system is responsible for global control flow and for sending out Remote Procedure Calls (\RemProc s) to direct workers to take an action. Workers are responsible for housing program data and \RemProc\ definitions. There can be several types of workers. Each type of worker can have a different \RemProc\ definition but all worker types must have a definition for each \RemProc\ in a program, even if it is trivial. A \RemProc\ contains a signal indicating which procedure to run and necessary arguments to accomplish the procedure. The control system may also contain global values that are common among workers. \RemProc s may direct workers to take any valid action regardless of complexity and may include worker-to-worker communication and worker-to-control communication. A program consists of one or more \RemProc s in a defined sequence. It is the controllers responsibility to house and manage this sequence to enact a program. This \virtMachine\ is completely hardware agnostic and is simply a division of roles and responsibilities. It is up to the compiler to decide hardware actions that enact this \virtMachine\ concept. For instance, it is possible (and already implemented in MACH) for a CPU to take on the roles of both the controller and all types of workers. We utilize this frequently to verify complex programs with small data sets before moving to large data sets and running on the \wse.

\begin{table}[h!]
    \centering
    \caption{Virtual Machine Roles and Responsibilities}
    \label{tab:VM_Roles}

    \begin{tblr}{colspec = {Q[c,m] Q[c,m] Q[c,m]},
                 row{1} = {font=\bfseries, bg=gray!50},
                 row{odd[2]} = {bg=gray!25},
                 }
        \toprule
        \SetRow{}
        {Roles \& \\ Responsibilities}         &   Controller     & Worker(s) \\
        
        \midrule
        
        Purpose &   {Program Control Flow  \\ Global Data Manipulation} & Non-Global Data Manipulation\\

        Procedure Role   & {Make \RemProc s \\ Respond to Workers as Needed \\ Operate on Global Data} & {Take action when \RemProc s are made \\ Send Data to Controller or other Workers as Needed \\ Operate on Non-Global Data}\\

        Data Responsibilities        &   {Control Flow Definitions \\ \RemProc\ Sequence \\ \RemProc\ Arguments \\ Global Program Data} & {\RemProc\ Definitions \\ Non-Global Program Data}\\
        \bottomrule
    \end{tblr} 
\end{table} 

\section{Mapping the \virtMachine\ to the \wse}

\begin{figure}[h]
    \centering
  \includegraphics[width=\textwidth]{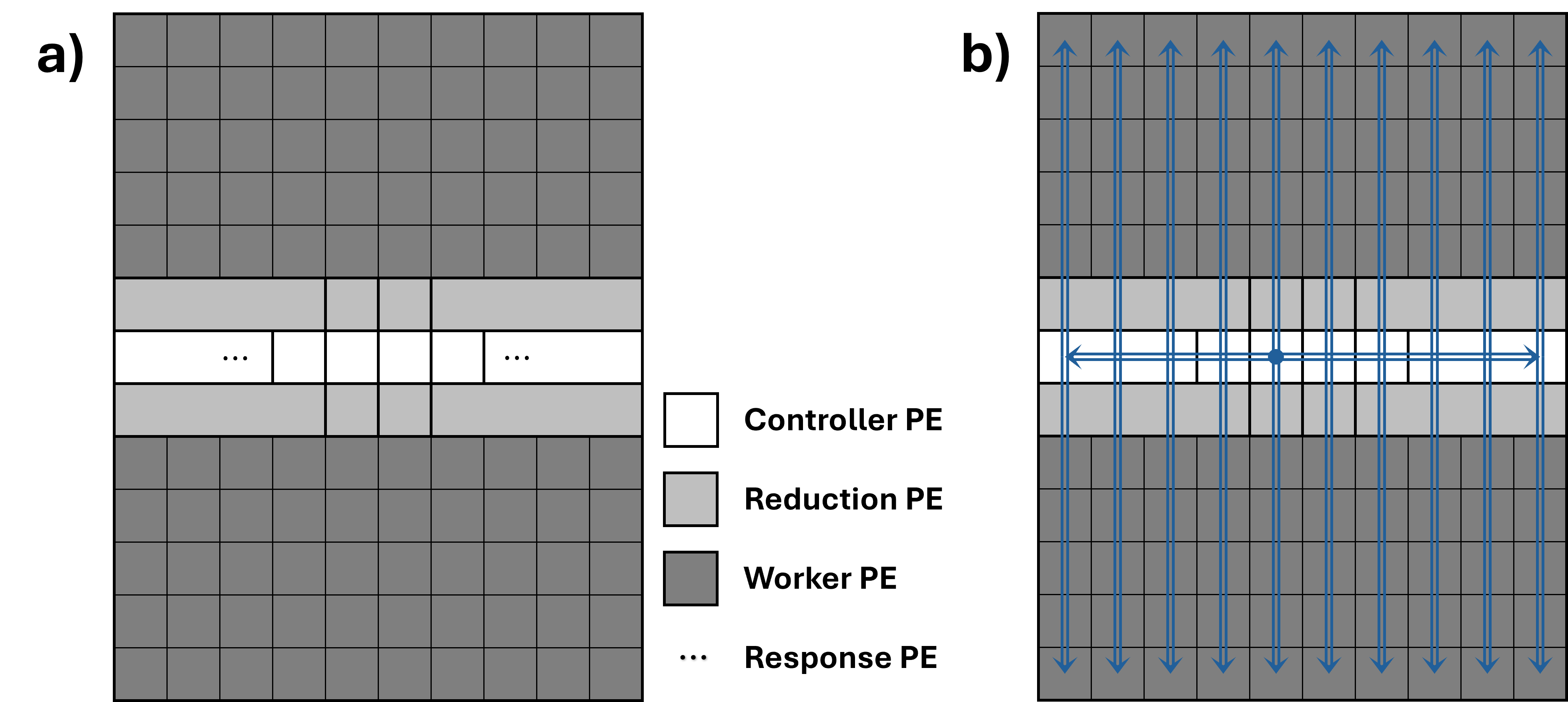}
    \caption{a) The Virtual Machine \ProcElem\ layout. Controller \ProcElem s are shown in white, Worker \ProcElem s in dark gray, and Reduction \ProcElem s in medium gray. The ellipsis denote an expandable set of Response \ProcElem s that are used to distribute program control flow data as needed.  b) The routing configuration for the control and arguments colors. Double-line arrows indicate a bus carrying multiple communication channels. The dot represents the source PE; in this case, every tile receives and transmits along the route, except for the strip of Controller PEs which only passes data through.}  
    \label{fig:Combined_DSL_Layout}
\end{figure}

We take a straightforward approach to assigning \ProcElem s to the \virtMachine\ roles. The choices made here are a series of compromises which we will discuss. The control system is placed on a row of \ProcElem s in the middle of a field of workers as shown in Fig. \ref{fig:Combined_DSL_Layout} labeled as \emph{Controller \ProcElem}. The control \ProcElem s hold all program control flow information. We define two kinds of workers within the worker field: \emph{Reduction \ProcElem s} and \emph{Worker \ProcElem s}. Worker \ProcElem s have the responsibility to house program data and participate in most tensor actions through \RemProc\ definitions. Reduction \ProcElem s participate only in reductions and do not house program data. 

The \wse\ has 24 virtual communication channels, referred to as \textit{colors}, a convention that simplifies diagrammatic representation using colored arrows. Fig. \ref{fig:Combined_DSL_Layout}(b) shows a blue bus of arguments and control colors being broadcast \RemProc s from the control system to the reduction and worker \ProcElem s. Reduction and worker \ProcElem s monitor this bus and execute \RemProc\ definitions as they are sent from the control system. The \virtMachine\ layout can be scaled to any size that is smaller than the \ProcElem\ array size on hardware. The control system is spread over a full row of \ProcElem s (hundreds of PEs depending on \wse\ version) resulting in a substantial program memory space.

Any \virtMachine\ mapping is a compromise, with benefits and drawbacks. We chose this \virtMachine\ mapping with a central control system and a set of worker and reduction \ProcElem s for several reasons. First, every \ProcElem\ must store both code and data. HPC applications often involve very large codebases - tens or hundreds of thousands of lines. However, they typically rely on a small subset of operations applied in long sequences. Thus, a controller/worker \RemProc\ system is attractive  as the worker field only needs to store a small subset of definitions rather than the full program. This leaves more memory available for program data, which is important. Second, the control system can span multiple \ProcElem s, enabling it to support the execution of  large programs. Third, the control system is centered in the \virtMachine\ layout because it minimizes mean distance between the control system and worker \ProcElem s. This is especially important when conducting global reductions that result in control flow action, which is common in linear solvers. Placing the control system in the middle of the \virtMachine\ has a drawback, though, as it increases latency to the chip edge, which adds some delay in communicating with off chip components. This does affect our plans for a runtime control environment. In most cases though, we use the x86 hosts for data collection from the wafer, where latency is less critical. Fourth, we have found that by vectorizing the control system and leveraging the high on-chip bandwidth, we can support fine-grained operations down to the level of single tensor operations with decent efficiency, especially for large tensors.

\subsection{The Control System}\label{subsec:control_system}
\begin{figure}[h]
    \centering
  \includegraphics[width=\textwidth]{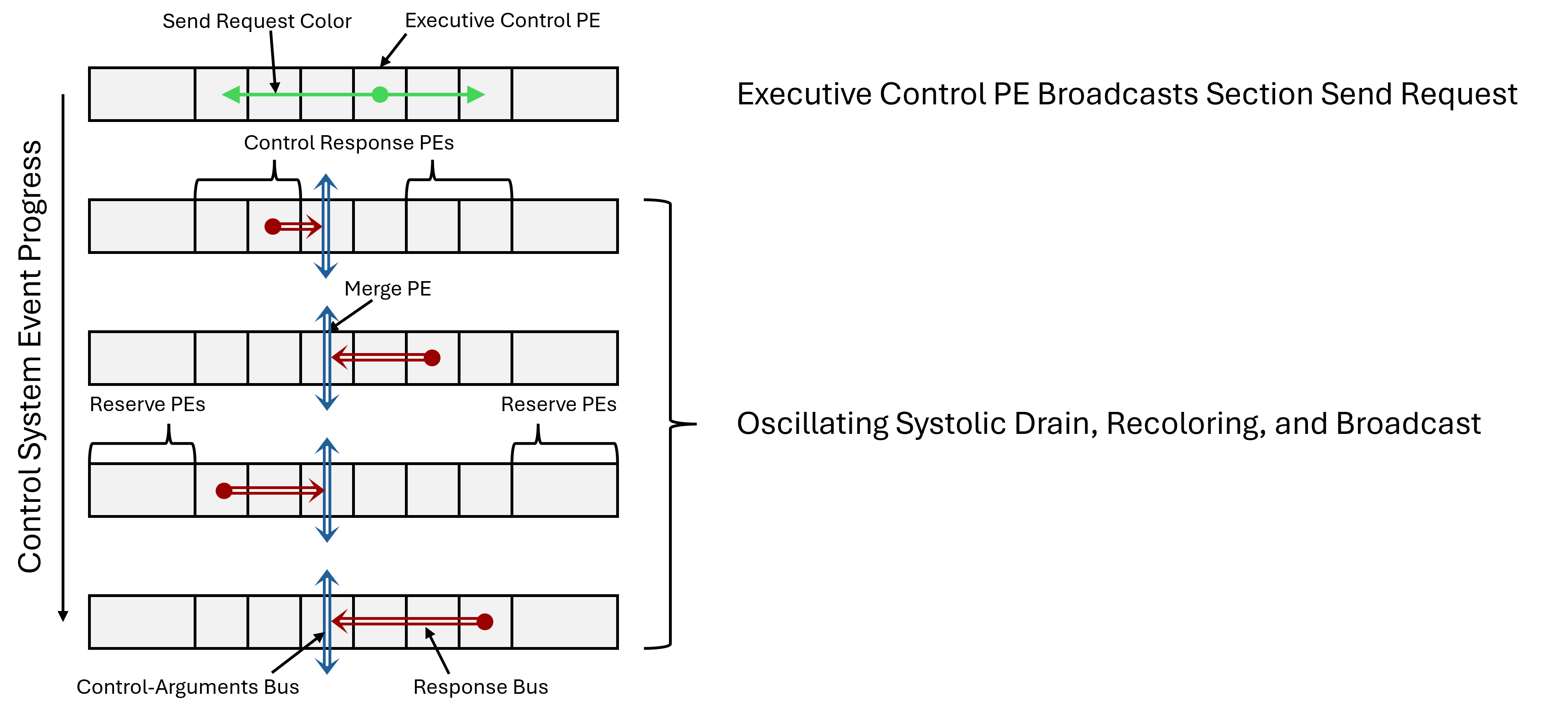}
    \caption{A detailed view of the distributed control system. In this case showing four \RespPE s with vector pieces for each control section. The \ExecPE\ broadcasts a message on a single color (green on top). Message arrival triggers \RespPE s to start filling send buffers with control and arguments data on the Response Bus. The \MergePE\ recolors arguments and control data to broadcast on the Control-Arguments Bus. It also transforms the control data on the Response Bus to control wavelets. The router is configured to sequentially drain the buffers on the \RespPE s in an oscillating systolic drain pattern.}  
    \label{fig:Distributed_Control}
\end{figure}

It is worth detailing the control system as it is a critical part of making the \virtMachine\ capable of running large programs with fine-grained kernels. Fig. \ref{fig:Distributed_Control} shows a detailed view of the Controller \ProcElem\ strip in the center of Fig. \ref{fig:Combined_DSL_Layout}. The distributed control strategy is shown at the granularity level of a single code section broadcast. The \ProcElem\ on right center of the \ProcElem\ layout is the Executive \ProcElem\ (\ExecPE) and contains all global program control flow statements (if, while, for, etc). Any code that exists between control flow statements gets compiled into a control vector and an arguments vector. These two vector sets are distributed as evenly as possible across the Response \ProcElem s (\RespPE s) adjacent to the center \ProcElem s. Fig. \ref{fig:Distributed_Control} shows four Response \RespPE s outside the center-left and center-right control \ProcElem s that contain distributed control/arguments vectors. If the program is large enough, the compiler can dedicate more \RespPE s to hold sequence data in the block marked with reserve \ProcElem\ s in Fig. \ref{fig:Distributed_Control} and ellipses in Fig. \ref{fig:Combined_DSL_Layout}(a). At full scale, there is as much as 48 MB of memory available in this strip of \ProcElem s. The \virtMachine\ mapping can expand to the full width of the wafer if necessary. 

The section broadcast is initiated by the \ExecPE\ which broadcasts a single data wavelet (an integer) that contains the section index on the \RespPE s. As soon as the value lands on the \RespPE s, the value is used to do an indirect lookup for the addresses of the arguments and control pieces assigned to \RespPE s by the compiler. Each \RespPE\ immediately begins sending arguments and control data towards the Merge \ProcElem\ (\MergePE) on two separate colors with an oscillating systolic drain pattern. A systolic drain is a pipelined flow where data moves in successive order from the closest \ProcElem\ to the farthest one from the exit. An oscillating systolic drain is useful when the exit is in the middle of a strip of \ProcElem s and data needs to be drained systolically from each side in progression from the inside out. The \MergePE\ is configured to recolor all incoming arguments onto the designated arguments color that worker and reduction \ProcElem s monitor. Similarly, control data (integers) are placed in a First-In-First-Out (FIFO) buffer and sent out as control wavelets on the control color. After sending its data, the \RespPE\ sends a router configuration wavelet that tells the \MergePE\ to reverse the direction of incoming data. When sent out, the wavelet also flips the outgoing router state on the \RespPE\ so that data flows from the next most outer \RespPE. When the outermost \RespPE\ sends its data, it sends a wavelet that both resets the routers of all inner \ProcElem s to start sending wavelets out and it triggers the \MergePE\ to switch direction again. 

The control system incurs about 10 cycles of latency between the time the \ExecPE\ broadcasts its initial signal to the time when wavelets begin to leave the \MergePE. This latency is a reasonable tradeoff,  given that we benefit from the ability to represent much larger programs on the wafer. In addition, the asynchronous, data-flow execution model of the \wse\ means that the \ExecPE\ can broadcast any sections that do not have a control flow statement dependent on a global reduction in very rapid succession. In these cases, the \RespPE s have data immediately available on completion of the last send and they act to do the indirect loads and fill the send buffers, so that they are ready to send as soon as the router is configured for outbound traffic. At worst, this system adds 10 cycles of latency to a global reduction,  which is negligible in benchmark tests. This latency occurs because the \ExecPE\ cannot continue execution until all previous commands have completed and broadcast their data. However, if the algorithm permits, the system can be configured to perform computation concurrently with the reduction. When the computational work is large enough, the entire reduction latency can be hidden.

It is important that the control system broadcasts control and arguments data as fast as possible which is why we chose to put the control and arguments sequence in a set of contiguous vectors on the control system. This allows the \ProcElem s containing the data to send it with minimal latency and instruction overhead. The \wse\ uses a reliable message system which ensures message wavelets are passed in a FIFO order and without corruption. A necessary consequence of the reliable communication system is that if messages cannot be consumed at the destination, message transmission must stop on the inbound side of each \ProcElem\ as message buffers fill. This proceeds in a cascade from destination to source and will eventually stop message transmission at the source (a process called back pressuring). In other words, if worker \ProcElem s cannot compute fast enough to consume the control/arguments signal, back pressure propagates to the controller \ProcElem s, causing the control system to slow down to the rate of the worker \ProcElem s. Ideally, the controller operates faster than the worker \ProcElem s and remains in a back pressured state as much as possible, ensuring the worker \ProcElem s are never idle waiting for instructions to operate on data.

We chose to implement a control system where the entire program representation resides on the control \ProcElem s because this action ensures that minimal latency is incurred during control flow operations. Minimizing latency of control signals is especially important for scientific codes that must make decisions on reduced values. The latency to host and back within the \wse\ architecture can be quite high in relative terms. Unlike other control systems that need an attached x86 host, our programs can execute without host interaction once they are compiled and loaded, if desired. However, when a user needs to save/load data, we configure hosts to act as workers that respond to the control system. In this way, hosts receive or send data according to the same computation graph executed on the \wse\ (see \S \ref{subsec:irg}). This configuration extends the asynchronous control strategy to a set of hosts which behave as though they were large memory, multi-core worker \ProcElem s on the wafer itself (although with higher latency, lower bandwidth access).

It is also worth noting that this control strategy enables fine-grained kernel support, with a 50-60 cycle setup cost before a typical operation begins. For most HPC codes, this is highly efficient, allowing scaling down to as few as 50 elements per processor (compared to some 12-15k elements per processor on traditional hardware). We find that we gain sufficient performance if all kernels are written at individual unary/binary operations, provided that we design software to fill memory with data and do not attempt to strong scale to smaller than about 50 elements per operation. If frequent, ultra-fine-grained data manipulation is required, it may be best to write custom kernels that offset the fixed setup cost, which is not difficult to do in our compiler system. We are also exploring kernel fusion techniques based on user definition and the compositional structure of the Intermediate Representation Graph (\IntRepGraph) discussed in \S \ref{subsec:irg}.

\subsection{The Reduction System}\label{subsection:reductionSystem}
\begin{figure}[h]
    \centering
  \includegraphics[width=\textwidth]{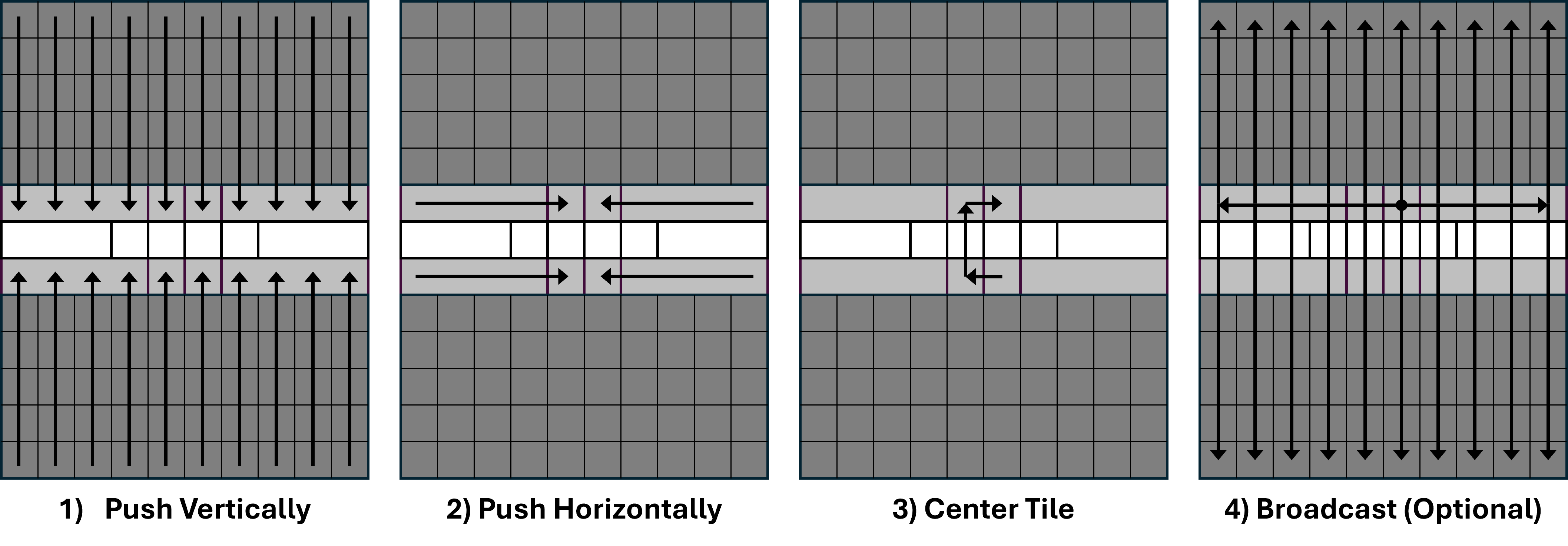}
    \caption{Stages of reduction on the \virtMachine. First, all worker \ProcElem s push data vertically to the reduction strip where it is accumulated in parts. Second, data is pushed horizontally where it is accumulated in four parts. Third, data is pushed to a single central tile where accumulation is finished. Lastly, it is optionally broadcast back to the worker \ProcElem s if the result is a \ULSclr\  or sent to the \ExecPE\ if it is a \GSclr.}  
    \label{fig:Global_Reduction}
\end{figure}

We dedicate two rows of \ProcElem s to reductions in the \virtMachine mapping. To accomplish a reduction, worker \ProcElem s push data to the reduction strips in a systolic drain, enabling efficient accumulation. The reduction \ProcElem s accumulate values and then push their data to the middle in a second systolic drain where the four inner \ProcElem s accumulate the values from the reduction strip. Finally, the four inner tiles push their data to a single tile in a third systolic drain that results in a globally accumulated value. The final value is either sent back to the worker \ProcElem s or is sent to the \ExecPE\ depending on the data structure specified in user code. Fig. \ref{fig:Global_Reduction} shows the stages of reduction on the \virtMachine mapping.

This system has some advantages for scalar reduction. In particular, worker \ProcElem s are completely free to do other tasks once the scalar is pushed to the router buffer. All processing and memory for accumulation is mutually exclusive to the worker \ProcElem s which free them to do overlapping functions if possible. This approach is well-suited for scalar reductions, which are typically latency-bound rather than limited by \NetOnChp{} bandwidth. As a result, this is desirable for pipelined linear solvers.

Tensor reduction is another matter. Depending on tensor size, operations can easily become bandwidth-constrained. We are currently exploring alternative tensor reduction options that may either necessitate further changes in the \virtMachine\ mapping or come with trade-offs. One promising option is based on a new in router reduction technique on the CS-3. The router supports mixed precision chained addition. With this methodology, a chain reduction from the outside inward could alleviate bandwidth constraints. However, with our current \virtMachine\ layout, we would probably have to copy data to a FIFO-oriented data structure (ie. a queue) and dedicate a microthread to send data within the reduction task, as the outside members are the last to receive the command from the \ExecPE, but the first to begin chain transmission. Additionally, since the input from the PE is half precision while the sum is single precision, care must be taken to conduct sums in single precision. For dot products and sums, this technique may be preferable. From an HPC perspective, it would be desirable to extend the in-router reduction capability to support all data types, and to include operations such as a maximum or minimum.

Tree, Chain, Two-Phase, and Auto-Gen reductions can be implemented on our \virtMachine\ mapping \cite{autogenReduction}. However, these techniques may not lend themselves well to overlapping communication and computation as they rely on parts of the worker field to participate in multiple stages of reduction which has an effect similar to blocking when algorithms depend on those neighbors for computation in a desired overlap. This effect is largely absent in the current scalar reduction, and would be slightly more pronounced in the in-router chain reduction, but at least the amount of work is balanced across the worker \ProcElem s which gives the asynchronous execution pathways a chance to balance as the sends progress through the worker field. While an uneven workload in the worker field is generally undesirable, it may be justified by gains in reduction efficiency, communication patterns, or other algorithm constraints.

\subsection{Moat Free}
We have transitioned to a moat free \virtMachine\ implimentation so that we can gain use of \ProcElem s that would otherwise be largely non-participatory in computation. This marks a substantial change from our previous \virtMachine\ implementation \cite{disruptiveChanges, recordIsing}. To support this change, we developed a participation filter in the \DomSpecLang\ to activate a subset of \ProcElem s within the worker field based on the user-defined slicing of the parent tensor. Aside from allowing the activation of a subset of workers, valid tensor slicing naturally prevents data from being shipped off the edge of the fabric making moats unnecessary. This comes at a very slight increase in kernel setup costs, an increased number of int16 values to place on workers, an extra value (or two) to transmit as an argument. Our compiler tracks unique slicing and creates masks to use in kernels. If necessary, we can pack the masks into a single bit should a large number be needed. Overall, this moat free approach is a worthwhile tradeoff to gain extra worker \ProcElem s and to add significant flexibility in the \virtMachine\ mapping.

\section{The Domain Specific Language}\label{sec:dsl_mapping}

We have developed a Domain Specific Language (\DomSpecLang) tailored to the \virtMachine\ described in \S \ref{sec:virtual_machine}. The \DomSpecLang\ is composed of a set of Object Oriented Data Structures (\oods s), an Intermediate Language (\IntLang), an Intermediate Representation Graph (\IntRepGraph), and a front end compiler that translates the high-level language to the \IntLang. \oods s are used to place and distribute data across the \virtMachine\ and provide operation definitions to act on data. The \IntLang\ is designed to be compatible with the \oods s and allow flexible data access patterns, similar to common tensor access patterns in high-level scientific computing languages. The methods in the \IntLang\ create a standardized \IntRepGraph\ which we can optimize and compile to the \wse.

\subsection{Object Oriented Data Structures}\label{OODS}

Our compiler must determine how to partition data across the spatial architecture,  identify which \ProcElem s hold  the data, and how to address and operate on the data. To accomplish this, we define several \oods s. Objects that have a singular value across the entire system are called Global Scalars (\GSclr s) and the memory is allocated on the \ExecPE. Similarly, rank 1 or higher tensors with common values across the system are called Global Arrays (\GArr s). \GArr s are also allocated on the \ExecPE\ (though we may distribute them across \RespPE s). Objects that constitute a scalar on each worker \ProcElem\ are called Local Scalars (\LSclr s) and are allocated on the worker \ProcElem s. Rank 3 or higher tensors are called Local Arrays (\LArr s) and are also allocated on the worker \ProcElem s. \LSclr\ objects can be configured as a Uniform-LS (\ULSclr) which always have the same value on all worker \ProcElem s. We implemented \ULSclr\ data structures since they enable the control system to vectorize references to memory as address values, resulting in the value being immediate on the control system and eliminating the need for indirect lookups. An indirect lookup on the executive control PE breaks up vectorization in the control system in the same way that a control flow event does. A \ULSclr\ thus functions as a GS but has characteristics that allow the control system to be fast. In addition, we employ a bank-aware memory placement system for \LArr s. This leaves holes to fill with \LSclr\ and \ULSclr\ values such that there is little-to-no impact on data representation capacity unless a large number of \LSclr/\ULSclr\ values are defined.

Each instance of a data structure represents a contiguous section of memory on the associated \ProcElem(s). The memory manager allocates memory at the same address and with the same extents on all associated \ProcElem s (see \S \ref{sec:mem_manager}). This design simplifies operations: only a few address values and a size need to be broadcast to specify the requisite data for an operation. In addition, each object contains operation overloads that represent tensor operations. These overloads  are used to  construct a static Intermediate Representation Graph (\IntRepGraph), which is later used for memory management and program optimization.

A natural data mapping emerges from the \virtMachine\ for dense tensors. In its simplest form, a two-dimensional tensor can be shared across all \ProcElem s in the form of a \LSclr\  where each worker \ProcElem\  holds a single scalar value. Similarly, three-dimensional arrays can be mapped such that each \ProcElem\  holds one axis in a vector. Since the \wse\ supports up to rank 4 tensors in its data descriptors, the same simple data mapping can be used to map up to rank 6 tensors provided enough memory exists on each \ProcElem\ for the dimensions held in memory.

Beyond this, a variety of alternative data mappings can be expressed within this \DomSpecLang\ while preserving the requisite data-processor locality needed for efficient operation. For instance, one or more of the dense dimensions can be mapped to multiple \ProcElem s in a nearby group rather than a single PE. Lastly, we are actively researching methods to map unstructured grids to this \DomSpecLang\ concept while maintaining processor-data locality - this will be addressed in future work.

\UseRawInputEncoding
\subsection{NumPy: The Compiler Source}
As our initial source language, we selected NumPy/Python due to its popularity, accessible Abstract Syntax Tree (\AbsSynTree), prevalent usage within our organization, strong documentation, first-class tensor support, and our personal familiarity. We also expect to be able to apply similar approaches to processor-centric intermediate representations, e.g., from LLVM-IR \cite{LLVM}, that would allow us to compile C++ or FORTRAN in a similar way. 

We strive to create a user experience that closely mirrors traditional NumPy and will function identically. In addition, we want the same source to be able to run on an x86 machine for rapid development and testing, then seamlessly compile and run on the \wse\ platform
 with consistent results. Other than a few minor concepts, we do not 
want the user to have to be an expert in spatial architectures to be 
able to write effective code. Several high-level, \wse\ compatible NumPy programs can be found on our GitLab
 repository. To date, we have developed programs that span a wide range 
of fields: several versions of Computational Fluid Dynamics, several 
versions of the Ising model, a two-phase subsurface model, an early 
Molecular Dynamics code, and an early Monte Carlo materials model. All 
of these models are built on a structured grid/lattice. We are also 
actively researching and developing comprehensive methods to support 
unstructured mesh methods.

For illustration, a small example NumPy program is shown in Listing \ref{exNPProg}. In this example, we import a specially maintained version of NumPy (line 2), the front-end \oods\ classes that are built around a dense tensor mapping (lines 3-6), and context managers (\contextMan) that direct the compiler to take specific actions within the body (line 7). The \texttt{Run\_On\_Host} \contextMan\ directs the compiler to run the code in the body at compile time without any lowering to the \IntLang. This is useful for data preparation with external Python packages. The \texttt{MACH\_Compiler\_Ignore} \contextMan\ directs the compiler to remove the statements from the \IntLang\ program. This makes code in these blocks execute only in the NumPy
 front-end execution, which is useful for post-analysis and data 
visualization with libraries that are not supported within our compiler 
environment. In this simple example, several arrays are declared (lines 
11-13). The data is used to declare and initialize some \oods\ variables (lines 16-18). We then loop over the elements in \texttt{myGA}
 (line 21). A running sum of the elements is computed on line 23. A 
conditional break is executed if the sum is greater than 100.0 on lines 
26-27. A tensor addition is performed on line 30. After the loop is 
complete, a reduce sum is performed to produce a \ULSclr\ result on line 32. Since the operation produces a new object that is not a direct instantiation from one of the \oods\ classes, type hinting as a \texttt{MACHScalar} is required.\newline

\begin{lstlisting}[language=Python, caption=Example NumPy source program, label=exNPProg, style=pythonstyle]
# Import our MACH NumPy, OODS classes, and context managers
import MACH.MACH_Numpy as np
from MACH.MACH_Numpy_Tools import MACH3DArray as M3a
from MACH.MACH_Numpy_Tools import MACH2DArray as M2a
from MACH.MACH_Numpy_Tools import MACH1DArray as M1a
from MACH.MACH_Numpy_Tools import MACHScalar as MS
from MACH.MACH_Numpy_Tools import Run_On_Host, MACH_Compiler_Ignore as Ignore

# Initialize normal numpy arrays on host
with Run_On_Host():
    nx, ny, nz = 10, 10, 10
    myLA_np = np.random.random((nx, ny, nz)).astype('float32')
    myGA_np = np.random.random((nz,)).astype('float32')
    
# Declare and initialize instances of OODS classes
myLA = M3a(myLA_np)
myGA = M1a(myGA_np)
myGS = MS(0.0,controller=True)

# Loop over elements in myGA
for gs in myGA[0]:
    # Keep a running sum of the elements
    myGS[0] = myGS[0] + gs[0]

    # Break if running sum is over 100.0
    if myGS[0] > 100.0:
        break

    # Do a tensor addition
    myLA[1:4, 3:5, 1:2] += myLA[1:4, 3:5, 8:9]

mySum:MS = np.sum(myLA[:,:,:])

# Take actions ignored by compiler
with Ignore():
    import matplotlib.pyplot as plt
    plt.imshow(myLA.view(np.ndarray)[2, :, :])
    plt.show()
\end{lstlisting}

Provided that a user installs our Python libraries, this code is executable on any x86 machine as normal python code (ie without compiling for the WSE hardware). When executed, \oods\ objects are treated as NumPy arrays and the program is executed in the host Python environment just like any other NumPy program.

\subsection{Compiling to the Intermediate Language}
It is necessary to translate the NumPy source into a form that enables us to create a standardized representation. To do this we use a custom built compiler that interacts with the source through the \AbsSynTree\ representation of Python using the \texttt{NodeVisitor} and \texttt{NodeTransformer} classes within the standard Python \texttt{ast} package. The purpose of this step is to translate source code into a form which can be used to create a standardized \IntRepGraph.

This step would likely be easier with strongly typed languages.
Because of Python's dynamic type system, no type information is stored in the \AbsSynTree\ which makes it difficult to determine whether objects are \oods\ types at compile time. 
To address this, we define two custom class decorators and a custom function decorator that utilize Python's existing type hinting system to provide strict type checking to a decorated classes and methods:
\begin{itemize}
    \item The first class decorator notifies the \AbsSynTree\ transpiler that the decorated class contains \oods\ variables that should be represented in our \IntLang. 
    When the \AbsSynTree\ transpiler encounters a class using this decorator, it automatically applies the other two decorators as appropriate.
    \item The second class decorator prevents silent retyping of class member variables, raising an error if a type mismatch is detected between an existing member variable and the variable being assigned. 
    \item The function decorator enforces type checking on method arguments using supplied Python type hint annotations. 
    If there is a mismatch of a supplied argument's type and the associated annotation, an error is raised. 
\end{itemize}

The \IntLang\ is written in Python and designed to be similar to languages with tensor operations as first-class citizens. The language is designed to closely follow NumPy, but provides the information needed to decompose tensors onto the \virtMachine , coordinate control flow and coordinate actions between wafer(s) and host(s). \newline

\begin{lstlisting}[language=Python, label=exILProg, caption=Example \IntLang\ program, style=pythonstyle]
# import our special veriosn of NumPy
import MACH.MACH_Numpy as np
# import OODS classes and on device for loop
from MACH.MACH_Local_Scalar import MACH_Local_Scalar
from MACH.MACHGlobal_Scalar import MACH_Global_Scalar
from MACH.MACH_Global_Array import MACH_Global_Array
from MACH.MACH_Numpy_Tools import Run_On_Host
from MACH.MACH_Loops import MACH_For_Loop

# Initialize normal numpy arrays on host
with Run_On_Host():
    nx, ny, nz = 10, 10, 10
    myLA_np = np.random.random((nx, ny, nz)).astype('float32')
    myGA_np = np.random.random((nz,)).astype('float32')
    

# Declare and initialize instances of OODS classes
myLA = MACH_Array(name='myLA', initData=myLA_np)
myGA = MACH_Global_Array(name='my_global_array', initData=myGA_np)
myGS = MACH_Global_Scalar(name='my_scalar', value=0.0, controller=True)

# define a loop on device the size of myG
with MACH_For_Loop('fl_0', myGA[:], controller=True) as fl_0:
    # Keep a running sum of the elements
    myGS[0] = myGS[0] + fl_0[0]

    # Break if running sum is over 100.0
    fl_0.conditional_exit('arg0 > arg1', [myGS[0], 100.0])

    # Do a tensor addition
    myLA.apply_length_filter([1, 4, None], [3, 5, None])[1:2, 0, 0] += \ 
       myLA.apply_length_filter([1, 4, None], [3, 5, None])[8:9, 0, 0]

mySum: MS = np.sum(myLA.apply_length_filter([None, None, None],
                                            [None, None, None])[:, 0, 0])
\end{lstlisting}

The compiler transforms the program in Listing \ref{exNPProg} into the \IntLang\ form shown in Listing \ref{exILProg}. The compiler converts the import statements to import the backend \IntLang\ \oods\ classes and adds a for loop context manager (lines 4-8). Front end \oods\ instantiations are transformed to \IntLang\ instantiations (lines 18-20). Since the iterable object in the for loop is an \oods\ object on the controller, the \AbsSynTree\ compiler converts the loop to a \contextMan . Converting the loop in this way places the control-flow on the \ExecPE\ and the body is executed within a for loop on the WSE. If a Python object is used as the iterator, the loop is not converted which has the same effect as unrolling a loop. Any references to the loop iteration object (\texttt{gs}) are converted to sliced instances of the for loop \contextMan\ (\texttt{fl\_0[0]}). The \texttt{\_\_getitem\_\_} dunder method of the for loop \contextMan\ is overloaded to support further lowering. The conditional break statement is converted to an attribute method call with a generic logic statement and list of arguments (line 28). This format is amenable to further lowering. 

One of the most important conversions is the conversion of \oods\ objects. The \IntLang\ \oods\ classes have an \texttt{apply\_length\_filter} method that we use to set participation within the worker \ProcElem\ array. This enables us to activate a subset of \ProcElem s to have finer grained control over the worker field. The arguments to this method are slices of \ProcElem s (start, stop, step) that mimic the slicing in a dense tensor of rank 2 or higher. The slicing provided in the arguments gets mapped to \ProcElem\ memory and is used in the participation filter of the standard kernel structure shown in Listing \ref{kernelExample}. In the simple dense mapping, we map the first two dimensions to the worker \ProcElem\ dimensions. Any remaining dimensions are mapped to the slice after the \texttt{apply\_length\_filter} method. The compiler does a similar transformation to \texttt{myLA} in the reduction on line 34 but converts the slicing to the entire range in all dimensions. The compiler does not need to convert \texttt{sum} because it is overloaded in \texttt{MACH\_Numpy} to handle \oods\ instances as well as normal \texttt{ndarray}s.

\subsection{The Intermediate Representation Graph}\label{subsec:irg}
By executing code written in our \IntLang, an \IntRepGraph\ is constructed, reflecting the data and order of operations present in the source. The resulting \IntRepGraph\ represents program control-flow, with the graph direction oriented from the start to the end of the program. Nodes in the graph describe operations and destination data access specifics, whereas edges describe source data access specifics. Each node is assigned a unique id reflecting execution order. Thus, evaluating nodes in order by id ensures dependencies are calculated in a just-in-time fashion.

\begin{figure}[h!]
    \centering
  \includegraphics[width=0.75\textwidth]{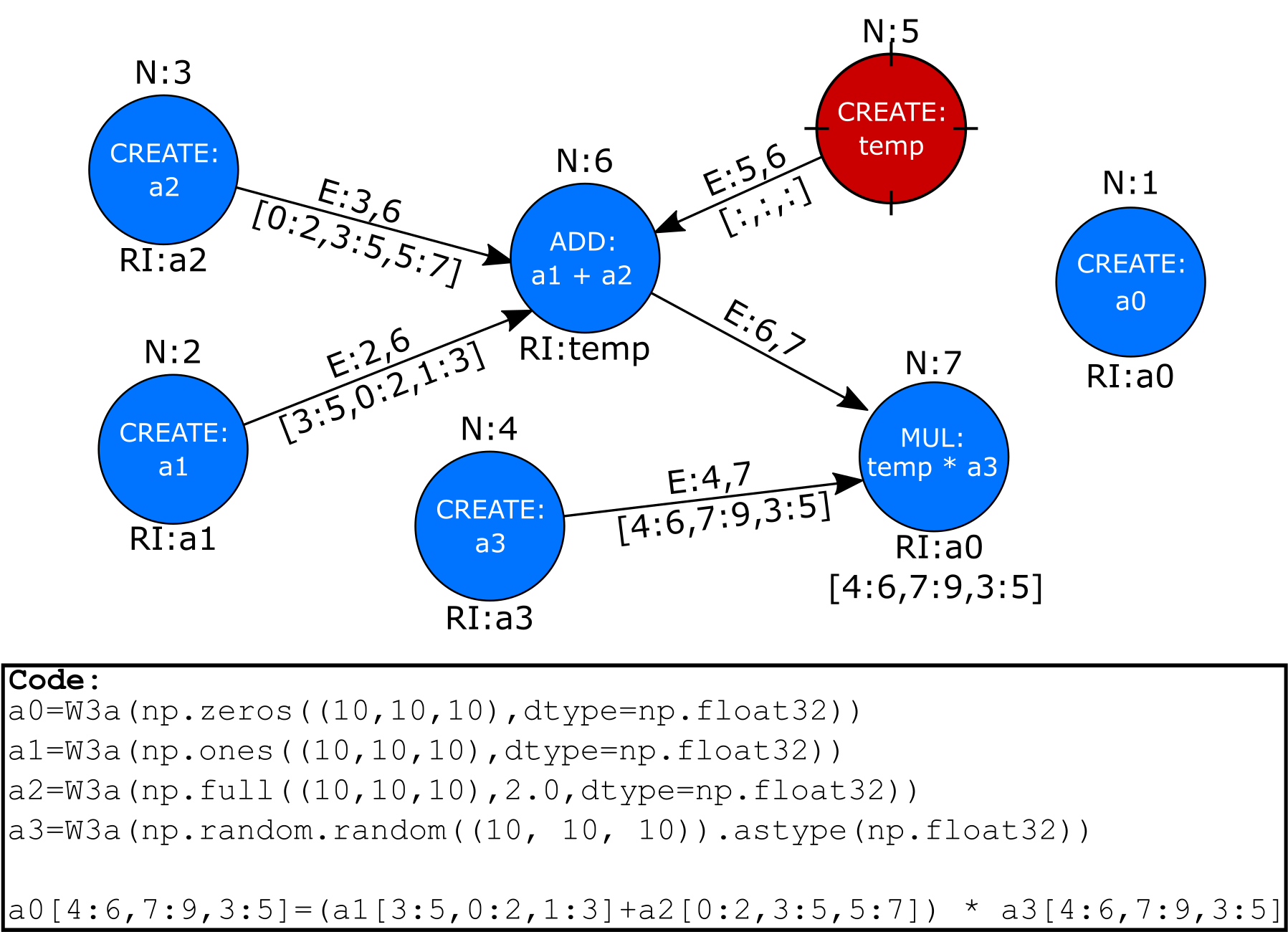}
    \caption{Simple \IntRepGraph\ representing combined array add and multiply operations, as presented in the \textbf{Code} box. Blue bubbles represent explicit operation nodes, while the spiky red bubble represents a compiler-created temporary node. Each node has an operation name and one or more address identifiers declared as operands. N:... denotes the unique node id, as well as the order of execution. RI:... refers to the result index, or the address identifier designating where the results of an operation are captured, optionally with a slice specified. Black arrows are directed graph edges that represent source data access, and are labeled with \textbf{E: ...,...}. The RI address identifier of an edge's source node is assigned as an argument to the operation represented by the destination node, with the slice below the edge label being applied.}  
    \label{fig:add_graph}
\end{figure}

An operation node is tagged with a unique identifier corresponding to a memory location and will be assigned an explicit memory address by the Memory Manager on compilation (see \S \ref{sec:mem_manager}). 
A simple, compound-operation graph example can be seen in Fig. \ref{fig:add_graph}.
We maintain a graph-in-graph hierarchy for sub-scopes. For instance, when a control-flow statement such as a \texttt{for\_loop} is encountered, a special node is placed into the parent graph, and the sub-scope body encapsulated in a sub-graph. Variables can be referenced between sub-graphs through the use of \texttt{subgraph\_transfer} operations. Sub-graphs are able to reference the same unique memory location identifiers, with the Memory Manager's liveness analysis, ensuring that any such memory locations will remain valid throughout the subgraph's evaluation.

\subsection{The Memory Manager}\label{sec:mem_manager}

The target language (Tungsten, \S \ref{sec:tungsten_target}),  follows a static, global memory model. Our compiler reserves a block of memory at compile time within which each program variable must be statically mapped to a specific address.
With only 48KB of memory available to store the combined program instructions and variables, it is prudent to reuse memory whenever possible.

The Memory Manager is responsible for enabling a smart use of limited available memory. 
The manager tracks active memory locations and captures the overall memory footprint for both the control and worker \ProcElem\ memory spaces.
This is done through a liveness analysis \cite{alfred2007compilers}, which follows the control flow established by the \IntRepGraph, evaluating each node in order. 
The Memory Manager emulates malloc/free-style operations to capture when and where memory is needed, marking emulated memory blocks as either \textit{allocated} or \textit{free}.
Each allocated block is tracked via an offset and a lifespan for the associated value, collectively referred to as an address entry.
Freed blocks of previously allocated memory are tracked and made available for reuse.
The sizes assigned to distinct values are determined by the variable size and precision. 
Blocks are aligned to any requested bank boundary and, when necessary, ensuring alignment along even 16-bit word boundaries.

The life spans of address entries are determined following patterns often utilized by smart pointers.
This is accomplished by the following rules: 
\begin{enumerate}
\item Persistent address entries provided with initialization data are allocated from the beginning of the program's execution. All others are allocated during the first operation that references them.
\item Persistent address entries explicitly designated as "output" values are treated as allocated throughout the duration of the program, allowing for their retrieval after the program's conclusion. All others are  marked as free after the last operation which references them.
\item Temporary address entries are always marked as allocated when they are first referenced in an operation, and marked as free after the final reference from a node.
\end{enumerate}
Applying the aforementioned rules creates opportunities for memory reuse, as life spans which do not overlap temporally can reference overlapping memory blocks without interfering with one another.
Thus, when a new allocated block is requested, the Memory Manager will attempt to find an existing free memory block to reassign, provided that it can satisfy the size requirements.
If multiple free blocks satisfy the size requirements, a \textit{best-fit} approach is used to choose from the candidates.
If a free block of sufficient size is found, it is bifurcated into an allocated block of the requested size and a free block encapsulating any remaining memory.
If no free block is large enough to be repurposed, then a new allocated block is appended to the list, increasing the overall memory footprint.

Once the analysis concludes, the largest memory footprint encountered during \IntRepGraph\ evaluation determines the size of the reserved memory space in Tungsten for storing program variables. Following this, the Memory Manager assigns memory addresses to all relevant program variables. The memory addresses values are used to form the arguments vectors that are distributed across the \RespPE s and also used to establish symbols in Tungsten files for persistent variables.

\section{Compiling for the \wse}\label{sec:tungsten_target}

On the WSE, our \virtMachine\ is implemented in Tungsten and Paint, which are proprietary languages developed by Cerebras for the \wse\ \cite{cerebrasMeeting, mdWSE}. Tungsten is a higher level language built around the Remote Procedure Call (\RemProc) concept and makes it simple to launch kernels and pass arguments between \ProcElem s on the \wse\ using the data flow concepts inherent in the hardware design. It also abstracts many of the complex instruction sets that would have to be managed at a lower level in a different language. Paint is a description language for the instantiation, layout, and interconnection of \ProcElem\ kernels written in Tungsten or other compatible languages. Our compiler must translate the \IntRepGraph\ to these languages in order to implement a program in accordance with our \virtMachine\ concept.

\subsection{Lowering to Tungsten}

Once optimized, the \IntRepGraph\ is lowered to several Tungsten programs that are compatible with the \virtMachine\ described in \S \ref{sec:virtual_machine}. Every program for the \virtMachine\ must have Tungsten definitions for the \ExecPE, \RespPE s, worker \ProcElem s, and reduction \ProcElem s. The \ExecPE\ and \RespPE s work together to maintain control flow and broadcast \RemProc s for the worker and reduction \ProcElem s. We now show the Tungsten output of the compiler for the program in Listing \ref{exNPProg}.

\subsubsection{Executive Processing Element Code Example}
The \ExecPE\ handles all global control flow statements and coordinates \RemProc s that are sent out on the control and arguments colors. \texttt{myGA} is a single precision global array with ten elements declared on line 16. The \ExecPE\ begins looping over the elements in the array on lines 22 and 24. The values in \texttt{myGA} are accumulated into a scalar (\texttt{myGs} on line 27). \newline

\begin{lstlisting}[label=controlerCodeExample, caption=Example Exectutive Control Kernel, style=cppstyle]
// declare communication sockts
xp socket ctrl_coord_color;

// declare unions 
union gs_sp {
    sp data;
    xp int16[2];
};

union gs_xp {
    xp data;
};

// declare memory blocks
xp var_start[n=26] address(0x5fe0);
sp myGA_ga[n=10] address(0x5fe0);
union gs_sp myGS_gs[n=1] address(0x5ff4);
union gs_sp temp_gs[n=1] address(0x5ff8);

{
    // loop over elements in myGa
    gfl_0_i ∈ [0, 10){
        // extract element into a temporary
        temp_gs.data = myGA_ga[fl_0_global_enum_gs.data];
    
        // accumulate value in myGS
        myGS_gs.data = myGS_gs.data + temp_gs.data;
    
        // conditionally exit
        sp myGS_c_exit[n=1];
        sp lit0;
        lit0 <- 100.0;
        myGS_c_exit[0] <- myGS_gs.data[0];
        if (myGS_c_exit[0] > lit0) break;
    
        // send section send request to direct tensor addition
        {
            ctrl_coord_color[] = wavelet(0);
        }
    }

    // send section send request to direct tensor reduction
    {
        ctrl_coord_color[] <- wavelet(1);
    }
}
\end{lstlisting}

The conditional exit is handled on lines 30-34. The \ExecPE\ signals to the \RespPE s to send section 0 over the \texttt{ctrl\_coord\_color} by calling the wavelet \RemProc\ on line 38 at the end of the loop, and signals the \RespPE s to send section 1 after the loops conclusion. See \S \ref{subsec:control_system} for a description of how these commands are distributed. This code is highly specific to each program and our compiler utilizes node-to-text translators to compose this program from the \IntRepGraph.

\subsubsection{Response Processing Element Code Example}
The \RespPE 's job is to listen for the \RemProc\ \texttt{wavelet} call and respond. We use a template with compiler variables and directives to compose this file as the sections on lines 7-17 and 21-24 are unique to each program. The same template is used for each \RespPE\ active in the program. The compiler divides data according to the description in Section \ref{subsec:control_system} so that all arguments and \RemProc\ calls are issued in the correct order. Listing \ref{rpeCodeExample} shows a full code sample from the first \RespPE. The code is identical on the second \RespPE\ except for the data section on lines 10-17 which is shown in Listing \ref{rpeCodeExample2}. \newline

\begin{lstlisting}[label=rpeCodeExample, caption=Response Processing Element Kernel, style=cppstyle]
// declare communication sockets
xp socket ctrl_coord_color;
xp socket ctrl_color_dist;
xp socket args_color_dist;

// define wavelet to flip router states after broadcasing data
wv const after_action_wavelet = {0x0, 0x0B40};

// define section data
const xp arguments_0[n=2] = {24546,24560};
const xp arguments_1[n=2] = {1,24544};
const xp send_arg_lengths[n=2] = {2,2};
const xp send_arg_address[n=2] = {0};
const wv ctrl_0[n=1] = {ctrl_color::ar_ar_addiii_float32_center};
const wv ctrl_1[n=1] = {ctrl_color::ar_reduceGlobalSumiii_float32_center};
const xp send_ctrl_lengths[n=2] = {2,2};
const xp send_ctrl_address[n=2] = {0};

program {
    // load the section addresses into memory
    send_arg_address[0] = &arguments_0;
    send_arg_address[1] = &arguments_1;
    send_ctrl_address[0] = &ctrl_0;
    send_ctrl_address[1] = &ctrl_1;

    // listing on ctrl_coord_color for wavelet
    forever main dispatch(ctrl_coord_color) {

        // define wavelet action
        wavelet(xp index) {
            // load the number of RPCs to send
            xp ctrl_length;
            ctrl_length = send_ctrl_lengths[index];

            // load the address to the RPC data to send
            xp ctrl_addy;
            ctrl_addy = send_ctrl_address[index];

            // load the number of arguments to send
            xp args_length;
            args_length = send_arg_lengths[index];

            // load the address of the arguments to send
            xp args_addy;
            args_addy = send_arg_address[index];

            // set the base address of the RPCs and arguments to send
            xp args[n=1];
            xp ctrl_sig[n=1];
            args.base = args_addy;
            ctrl_sig.base = ctrl_addy;

            // send the arguments and control data in parallel to the M-PE
            parallel {
                {
                    // send section arguments over args_color_dist
                    i in [0,args_length) args_color_dist[] <- args[i];
                    // send control wavelet to flip communication direction on
                    // the M-PE and reset downstream routers on other R-PEs
                    args_color_dist[] <- control(after_action_wavelet);
                    // flip my own router output direction if needed
                    args_color_dist.flip;
                }
                {
                    // send section arguments over ctrl_color_dist
                    j in [0,ctrl_length) ctrl_color_dist[] <- ctrl_sig[j];
                    // send control wavelet to flip communication direction on
                    // the M-PE and reset downstream routers on other R-PEs
                    ctrl_color_dist[] <- control(after_action_wavelet);
                    // flip my own router output direction if needed
                    ctrl_color_dist.flip;
                }
            }
        }
    }
}
\end{lstlisting}

The wavelet data construction on lines 14-15 is of special interest. We define \RemProc\ calls in this manner because the numeric integer value of the \RemProc\ is only determined by 
Tungsten during the build process. \texttt{ctrl\_color::ar\_ar\_addiii\_float32\_center} instructs the Tungsten compiler that we intend to use the \RemProc\ \texttt{ar\_ar\_addiii\_float32\_center} and to look for any definitions for it on \ProcElem s that are listening on \texttt{ctrl\_color}. Tungsten will replace this array of symbols with integer values compatible with its internal \RemProc\ system. 
 This system also allows us to define more \RemProc s than are supported by the hardware's task table. In such cases, Tungsten will create virtual tasks and a means to switch context.

When the \RespPE\ receives a \texttt{wavelet} call, it uses the \texttt{index} argument to indirectly load arguments and lengths and addresses for data to send (lines 33-45). It sends the data on their respective colors on lines 57 and 66. It sends the wavelet necessary to flip downstream routers on each color on 60 and 69 followed by an attempt to flip the output router state. In this example, flipping the state has no affect, but would allow a systolic drain if there were more than one \RespPE\ on each side. The \texttt{parallel} keyword launches these instructions on microthreads and synchronizes the microthreads before exiting the \RemProc.\newline

\begin{lstlisting}[label=rpeCodeExample2, caption=Example data section on second \RespPE, style=cppstyle,  firstnumber=9]
// define section data
const xp arguments_0[n=4] = {24546,1,1,0};
const xp arguments_1[n=4] = {0,10,24566,0};
const xp send_arg_lengths[n=2] = {4,4};
const xp send_arg_address[n=2] = {0};
const wv ctrl_1[n=1] = {ctrl_color::special_reduceBroadcast_float32};
const xp ctrl_0[n=1] = {0};
const xp send_ctrl_lengths[n=2] = {0,2};
const xp send_ctrl_address[n=2] = {0};
\end{lstlisting}

\subsubsection{The Merge Processing Element Example}
The \MergePE 's job is very simple. Every piece of data that comes in on \texttt{args\_color\_dist} is immediately sent out on \texttt{args\_color} (line 12 in Listing \ref{mergeCodeExample}). Any data coming in on \texttt{ctrl\_color\_dist} is read into a \texttt{FIFO} and sent out as a control wavelet on \texttt{ctrl\_color} (lines 13-14). This file is static for any program and is simply copied to the build directory. \newline

\begin{lstlisting}[label=mergeCodeExample, caption=Merge Processing Elemnet Kernel, style=cppstyle]
// declare communication sockets
wv socket ctrl_color;
xp socket args_color;
wv socket ctrl_color_dist;
xp socket args_color_dist;

// delcare a fifo with attached memory with 1000 element capacity
xp fifo_mem[n=1000];
wv fifo ff = {.mem=fifo_mem};

// recolor data
i in [0,Inf) args_color[] <- args_color_dist[];
i in [0,Inf) ff[] <- ctrl_color_dist[];
i in [0,Inf) ctrl_color[] <- control(ff[]);
\end{lstlisting}

\subsubsection{Worker Kernel Example}\label{subsec:kernel}
Worker \ProcElem 's are configured to listen for an \RemProc\ on \texttt{ctrl\_color} on line 9 of Listing \ref{kernelExample}. An \RemProc\ signal is a numeric value associated with a user-defined name like \texttt{ar\_ar\_addiii\_float32\_center}, \texttt{ar\_reduceGlobalSumiii\_float32\_center}, or \texttt{special\_reduceBroadcast\_float32}. When the \ProcElem\ receives the \RemProc\ signal, it executes the correspoinding code body. All \ProcElem s configured to listen to a color must have an \RemProc\ definition that matches the send-side signature. In the event that a particular \ProcElem\ is nonparticipatory in an \RemProc, it reads data from the arguments color and performs no other task. This action is important to prevent the  arguments channel from becoming clogged. 

For the addition kernel, we declare memory aliases for the operands and a spot to hold \RemProc\ arguments (lines 15-18). After the arguments are read into memory, the values are used to set the base addresses of the memory aliases for the operands (lines 24-30). Our compiler creates participation filters from the \IntRepGraph\ and stores them in local memory. The filter value is either 0 or 1. The last value loaded is the product of the participation filter and the length broadcast from the controller. This is a relatively efficient means to control participation in an operation within a group of \ProcElem s (lines 33-38). The last line is a single-instruction vector addition (line 41). In this way, this kernel performs a dense tensor addition on all or part of two dense tensors for which data is local to a \ProcElem.

The reduction kernel follows largely the same initial pattern until line 69 where an accumulator is declared and initialized. A local-to-\ProcElem\ accumulation is performed on line 73 and the value is transmitted onto \texttt{reduction\_1} in the conditional on lines 77-83 (Note: \texttt{reduction\_reset\_mask} is a geo-var that enables Tungsten to treat the conditional as a compiler directive, and only one of the branches is compiled based on the value of \texttt{reduction\_reset\_mask}). The scalar transmission is followed by either a reset wavelet or an output direction flip depending on the position in the systolic drain. The \texttt{special\_reduceBroadcast\_float32} kernel again follows a similar pattern of reading data from the arguments and setting an address. In this case, a single address value is read and used to store a scalar value from the \texttt{reduction\_broadcast} color on line 97. Separating the initial send in \texttt{ar\_reduceGlobalSumiii\_float32\_center} from the resulting receive  in \texttt{special\_reduceBroadcast\_float32} allows he worker \ProcElem s  to conduct work while the reduction is proceeding on the routers and the reduction \ProcElem s.\newline

\begin{lstlisting}[label=kernelExample, caption=Example Tungsten kernel, style=cppstyle]
// declare communication sockets
xp socket ctrl_color;
xp socket args_color;
xp socket reduction_1;
xp socket reduction_broadcast;
xp param reduction_reset_mask;

// configure the PE to listen for RPCs on the ctrl_color channel
forever main dispatch(ctrl_color){

    // define an array addition between float32 operands
    ar_ar_addiii_float32_center() {

        // declare local memory aliases
        sp s0[n=1];
        sp s1[n=1];
        sp dst[n=1];
        xp args_recv[n=5];

        // read arguments from arguments color
        i_recv in [0, 5) args_recv[i_recv] = args_color[];

        // set source and destination addresses
        xp temp;
        temp = args_recv[0];
        s0.base = temp;
        temp = args_recv[1];
        s1.base= temp;
        temp = args_recv[2];
        dst.base = temp;

        // read and configure participation filter
        xp len_mul;
        len_mul = args_recv[3];
        xp len_mul_load[n=1];
        len_mul_load.base = len_mul;
        xp length;
        length = args_recv[4] * len_mul_load[0];

        // do operation
        i in [0, length) dst[i] = s0[i] + s1[i];
    }

    // define a reduction kernel
    ar_reduceGlobalSumiii_float32_center() {

       // declare local memory aliases
       sp s0_sp_tsk[n=1];
       xp args_recv[n=4];

       // read arguments from arguments color
       i_recv in [0, 4) args_recv[i_recv] <- args_color[];
       
       // set source and destination addresses
       xp temp;
       temp = args_recv[1];
       s0_sp_tsk.base = temp;

       // read and configure participation filter
       xp len_mul;
       len_mul = args_recv[2];
       xp len_mul_load[n=1];
       len_mul = 24568 + len_mul;
       len_mul_load.base = len_mul;
       xp length;
       length = args_recv[3] * len_mul_load[0];

       // declare and initialize an accuulator
       sp accumulator;
       accumulator = 0.0;

       // locally reduce values on each worker PE
       i in [0, length)  accumulator <- accumulator + s0_sp_tsk[i];

       // send accumulator with output router flip or reset wavelet
       // depending on position in the systolic drain
       if (reduction_reset_mask == 1) {
           (sp)reduction_1[] <- accumulator;
           reduction_1[] <- control(reset);
       } else {
           (sp)reduction_1[] <- accumulator;
           reduction_1.flip;
       }

   }

   // define a reduction recieve kernel
   special_reduceBroadcast_float32() {

       // read the address from the args_color and set the result base
       xp temp;
       temp <- args_color[];
       sp result[n=1];
       result.base <- temp;

       // read the value from the reduction_broadcast color
       result[0] <- (sp)reduction_broadcast[];
   }
}
\end{lstlisting}

\subsubsection{Reduction Strip Example}

The reduction \ProcElem s do not participate in the addition or the receipt of the broadcast so they just have code that reads the arguments and dumps them into a harmless spot in memory. For the sake of brevity, we show the code from only the upper right reduction \ProcElem\ that participate in all the accumulation steps. The code for the other \ProcElem s in this group is similar but may contain a send or a reset wavelet to manage the systolic drains. The entire code is available in our GitLab repository \cite{mach_program}.

After reading the arguments and declaring an accumulator, the accumulator is initialized from the first value received from \texttt{reduction\_1} (line 37). The kernel then directs all \ProcElem s in the reduction strip to start sending data inward on \texttt{reduction\_2}. Each participating \ProcElem\ reads these values and accumulates them on line 43. In the case in Listing \ref{exNPProg}, the accumulation is going to result in a \ULSclr\ value, so the compiler sets the first argument to 1 so that \texttt{is\_local == 1} evaluates to true. In this case, the inner \ProcElem s send data on \texttt{reduction\_4} and it is accumulated on line 50 and sent out to the workers on \texttt{reduction\_broadcast} on line 52. If the user were to include \texttt{controller=True} among the arguments to \texttt{np.sum}, the compiler would set the first argument to 0, and the \ProcElem\ would direct the accumulation result to \texttt{reduction\_3}, which sends data to the \ExecPE\ for final accumulation.\newline

\begin{lstlisting}[label=reductionExample, caption=Example Reduction Strip Tungsten Kernel, style=cppstyle]
// declare communication sockets
xp socket ctrl_color;
xp socket args_color;
xp param n_half;
xp socket reduction_1;
xp socket reduction_2;
xp socket reduction_3;
xp socket reduction_4;
xp socket reduction_broadcast;
xp param center_reduction_pe;
xp param n_w_half;

wv const reset = {0x0,0x0180};

// configure the PE to listen for RPCs on the ctrl_color channel
forever main dispatch(ctrl_color){

   // define an array addition between float32 operands
   ar_ar_addiii_float32_center(){
       // read and dump arguments to keep pipeline from backpressuring
       xp dump[5];
       i_recv in [0, 5) dump[i_recv] <- args_color[];
   }

   // define a reduction kernel
   ar_reduceGlobalSumiii_float32_center() {
       // declare local memory aliases
       xp args_recv[n=4];

       // read arguments from arguments color
       i_recv in [0, 4) args_recv[i_recv] <- args_color[];
       xp temp;

       // declare accumulator
       sp accumulator;
       // initialize accumulator with the first value from the fabric
       accumulator <- (sp)reduction_1[];

       // do the accumulation with remaining H/2-1 elements (step 1)
       i in [0,n_half) accumulator <- accumulator + (sp)reduction_1[];

       // do accumulation from the reduction strip (step 2)
       i in [0,n_w_half) accumulator <- accumulator + (sp)reduction_2[];

       // read value if destination is ULS or GS
       xp is_local;
       is_local <- args_recv[0];
       if (is_local == 1) {
           // do accumulation from inner 4 neighbors (step 3)
           i in [0,3) accumulator <- accumulator + (sp)reduction_4[];
           // broadcast to workers
           (sp)reduction_broadcast[] <- accumulator;
       } else {
           // send data to the E-PE for final accumulation
           (sp)reduction_3[] <- accumulator;
           reduction_3.flip;
       }
   }

   // define a reduction receive kernel
   special_reduceBroadcast_float32(){
       // read and dump arguments to keep pipeline from backpressuring
       xp dump[n=1];
       i_recv in [0, 1) dump[i_recv] <- args_color[];
   }
}
\end{lstlisting}

\subsection{Lowering to Paint}
Paint is a convenient tool to "paint" properties on a region of \ProcElem s. Paintable properties include (but are not limited to) color routes for \ProcElem-to-\ProcElem\ communication, \ProcElem\ level code, geo-vars, and names. Our compiler builds this file dynamically using a mix of templates, 
compiler directives, and compiler variables. Optional features in the \virtMachine\ can
 be added and removed as needed based on the operations defined in a 
user's code. Rather than showing the file in its entirety, we now 
present selections of important concepts knowing that the full code is 
available in our repository \cite{mach_program}.\newline

\begin{lstlisting}[label=colorDeclaration, caption=Example Color Declaration, style=paint]
let rxColorData = color()
let txColorData = color()
let ctrl_color = color()
let args_color = color()
let ctrl_color_dist = color()
let args_color_dist = color()
let ctrl_coord_color = color()

let reduction_1 = color()
let reduction_2 = color()
let reduction_3 = color()
let reduction_4 = color()
let reduction_broadcast = color()
\end{lstlisting}

First, communication channels are declared as colors using the \texttt{let} keyword, as shown in Listing \ref{colorDeclaration}. Paint also allows function definitions that help create tessellations. We use this to create space filled regions with specific communication patterns and a variety of communication channels. An example is shown in Listing \ref{examplePaintFunction}. The \texttt{worker\_upper\_checker\_gen} function accepts two dimensions (\texttt{x} and \texttt{y}) and two colors (\texttt{c0} and \texttt{c1}). Two \texttt{tile}s (another name for \ProcElem s) are declared (\texttt{aa} and \texttt{bb}) and a geo-var (\texttt{@type}) is painted on the tiles. A geo-var is a regionally defined compiler directive that can be used in code as a parameter that directs compilation in conditionals and other logical flow. On lines 4-7, the communication patterns are painted in a defined sequence. An \textit{advance} wavelet command will change the communication pattern from state to state. \texttt{Ring} indicates that the state machine should return to its first state when advancing from the last. \texttt{C}, \texttt{R}, \texttt{L}, \texttt{U}, \texttt{D} indicate Center, Right, Left, Up, and Down relative to a \ProcElem s own position, respectively. \texttt{R>C} indicates that a \ProcElem\ is receiving data from the right while \texttt{C>L} indicates that a \ProcElem\ is sending data to the left. In their first state, \texttt{aa} is receiving data from the right on \texttt{c0} and sending data to the left on \texttt{c1} while \texttt{bb} is performing the same with the colors swapped. The \ProcElem s are stacked in an alternating checker pattern on line 8 and then the space is filled to the extents of x and y on lines 9-10. This creates a space filled checker pattern that allows nearest neighbors to simultaneously send and receive data using two different channels, which shift communication direction upon transmission an \textit{advance} wavelet. This is a key part of our \virtMachine\ design that enables efficient nearest-neighbor communication.\newline

\begin{lstlisting}[label=examplePaintFunction, caption=Example Paint function, style=paint]
define worker_upper_checker_gen(x y c0 c1) {
    let aa = tile() aa : @type=0
    let bb = tile() bb : @type=1
    aa : paint(c0 [R>C; D>C; L>C; U>C; ring])
    aa : paint(c1 [C>L; C>U; C>R; C>D; ring])
    bb : paint(c1 [R>C; D>C; L>C; U>C; ring])
    bb : paint(c0 [C>L; C>U; C>R; C>D; ring])
    let checker_pattern = vstack(hstack(aa bb) hstack(bb aa))
    checker_pattern : hstackrep(split(x 2) 'x_rep)
    checker_pattern : vstackrep(split(y 2) 'y_rep) }
\end{lstlisting}

We define several similar functions to fill \ProcElem\ space with our \virtMachine. \ProcElem\ code is painted onto a region with the \texttt{code} keyword, as shown in Listing \ref{exampleCodePaint}. In this example, a rectangular region of \texttt{W $\times$ H/2} \ProcElem s is declared on line 1. Two instances are stacked on top of one another with three rows of \ProcElem s between them on line 2. The \texttt{worker.w} Tungsten file is then painted to this region of \ProcElem s with several arguments being passed to the Tungsten file in a space-separated list on lines 3-11. Every \ProcElem\ in this region executes code compiled from the the same Tungsten source code.\newline

\begin{lstlisting}[label=exampleCodePaint, caption=Example Painting Code, style=paint]
let worker_code = rect(0 0 W H/2 'x 'y)
worker_code : vstackrep(2 3 'y_half)
worker_code : code('worker [ctrl_color:ctrl_color
                            args_color:args_color
                            c0_data:rxColorData
                            c1_data:txColorData
                            reduction_1:reduction_1
                            reduction_broadcast:reduction_broadcast
                            loopColorWorker:loopColorWorker
                            type:@type
                            reduction_reset_mask:@reduction_reset_mask])
\end{lstlisting}

We repeat this process to build up the \virtMachine\ using a combination of \texttt{union}, \texttt{stack}, and \texttt{place} operations. Unions are used to overlap regions and arbitrate their final properties. Stacks are used to place regions next to one another. Place operations, conversely, allow offset placement of a region.\newline

\begin{lstlisting}[label=exampleFinalUnion, caption=Final Construction of the VM, style=paint]
let VM = union(worker_code vstack(w_lower_ckr middle_ckr w_upper_ckr))
\end{lstlisting}

Construction of a moat-free \virtMachine\ is appreciably more complex and involves several function definitions and unions to enable checkerboard routing. However, this complexity is justified given the ability to use more \ProcElem s. At the end, all defined regions are joined with a union to create the \virtMachine, as shown in Listing \ref{exampleFinalUnion}. Here, \texttt{worker\_code} is the region containing the code property, \texttt{w\_lower\_ckr} and \texttt{w\_upper\_ckr} are regions with communication patterns and geo-vars painted on them using space filling functions like those in Listing \ref{examplePaintFunction}, and \texttt{middle\_ckr} is a composite region that contains all definitions for the control and reduction \ProcElem s.

\section{Enabling Technology on the WSE}\label{enablingTech}
We want to highlight a few particularly enabling technologies on the \wse\ and how we use them in our compiler to broaden our \DomSpecLang. Some of these concepts could be shared with other processors in the spatial architecture class.

\subsection{Local Memory Slicing}\label{localMemSlice}
In the simplest data mapping for our \DomSpecLang\, a dense three-dimensional tensor is mapped such that two dimensions span the \ProcElem\ dimensions and the third is held in memory. Operation scaling progresses in proportion to $N \times M \times L/N_p$. Here, $N$, $M$ and $L$ are the dimensions of a tensor and $N_p$ is the number of processors applied to do the operation. For the \wse\ and with a simple mapping, $N_p=N \times M$, thus scaling is proportional to $L$ provided that $N$ and $M$ do not exceed \ProcElem\ dimensions. The memory system and inherent hardware tensor support within \wse\ resulted in very good strong and weak scaling for simple structured grid, stencil problems \cite{disruptiveChanges, recordIsing, sai2023massively, sai2024matrix}.

While this is interesting, the structured grid restriction limits the application space for the \DomSpecLang. To overcome these restrictions, we take advantage of the individually programmable nature of each \ProcElem\ in the \wse. Namely, we allow dynamic slice-lengths for every \ProcElem\ in the worker field. We can express this with non standard NumPy syntax thanks to our specialized \texttt{numpy.ndarray} subclass and \AbsSynTree\ compiler. The following is a valid code in \texttt{MACH\_NumPy}.

\begin{lstfloat}
\begin{lstlisting}[language=Python, style=pythonstyle, label=nonUniformSlicing, caption=Example Non-uniform memory axis slicing]
import MACH.MACH_Numpy as np
from MACH.MACH_Numpy_Tools import MACH3DArray as M3a, MACH2DArray as M2a

stop:M2a = np.random.randint(0, 5, (10, 10))
s1:M3a = np.random.random((10,10,10))
s2:M3a = np.random.random((10,10,10))
dst:M3a = np.zeros_like(s2)

dst[:,:,:stop] = s1[:,:,:stop] + s2[:,:,:stop]
\end{lstlisting}
\end{lstfloat}

In this code, a $10 \times 10$ array of worker \ProcElem s is allocated. Each \ProcElem\ will first generate a \LSclr\ filled with random integer values in the $[0, 5)$ range. Next, two \LArr s are allocated and filled with random floats in the $[0, 1)$ range followed by another \LArr\ filled with zeros. An addition is then performed up to the first \texttt{stop} elements on each \ProcElem. The reference to \texttt{stop} is encoded as an indirect lookup performed within the memory space of each worker \ProcElem. The value retrieved for \texttt{stop} is then used to define the extent of elements for a given \ProcElem\ to be included in the tensor operation.

At first glance, this may seem like a trivial change. However, it is a key component to expanding the flexibility of our \DomSpecLang. This technology is used heavily in Lagrangian applications like molecular dynamics. In these applications, atoms are free to move throughout a domain spread across multiple \ProcElem s. Because of this capability, we do not have to globally track the maximum number of atoms on each processor to set loop sizes or use masked operations. The \ProcElem s themselves can be coded to track their own atom counts and use that count in their loops. It is also a key component in expanding our \DomSpecLang\ to support unstructured grids where cell, face, and neighbor counts are different across the worker field.

\subsection{Fused Gather/Scatter Loopback Operations}

The \wse\ also supports wavelet indexing and wavelet addressing, where a vector register can be set up to calculate its address based on data arriving on a predefined fabric color. This is a useful form of indirection. In particular, a \ProcElem\ can be set up to loop indices or addresses back to itself. In this configuration, a fused-gather-op or a fused-scatter-op can proceed at one operation every other cycle on average. Further, it is possible to make one of the sources or the destination a channel on the router. In combination with the Local Memory Slicing capability, we can write efficient \ProcElem\ code for unstructured grid applications within our \DomSpecLang. On the NumPy front end, we implement this with \texttt{take\_along\_axis} and \texttt{put\_along\_axis}. 

The interesting part lies in the kernel implementation on each \ProcElem. Instead of performing an indirect lookup in an explicit loop, each \ProcElem\ sends out a vector of indices to the router on a dedicated channel that is looped back to itself. The \ProcElem\ is configured to listen to this channel, and when data returns on this channel, it is used to modify the base pointer into a tensor operation. This is accomplished by declaring a loopback socket (line 4) and using the socket as an index in a memory alias (line 42). Listing \ref{gatherMulSend} shows a very useful operation. Here, two microthreads are used in parallel to send the index information back to itself and the gather is fused with a multiply and a send on line 40. An implementation like this allows the fused operator to proceed at an average of about two cycles per element. Such kernels are very useful for unstructured grid operations and can be very efficiently used for algorithms with indirection.\newline

\begin{lstlisting}[language=C++, style=cppstyle, label=gatherMulSend, caption=A fused gather-mul-send operation]
// declare communication sockets
xp socket ctrl_color;
xp socket args_color;
xp socket loop_color[loopback];
sp socket send_color;

// configure the PE to listen for RPCs on the ctrl_color channel
forever main dispatch(ctrl_color){
    // define an RPC kernel
    gather_mul_send() {
        // declare local memory aliases
        sp s0[n=1];
        sp s1[n=1];
        xp index[n=1];
        xp args_recv[n=5];

        // read arguments from arguments color
        i_recv in [0, 5) args_recv[i_recv] = args_color[];

        // set source and destination addresses
        xp temp;
        temp = args_recv[0];
        s0.base = temp;
        temp = args_recv[1];
        s1.base= temp;
        temp = args_recv[2];
        index.base = temp;

        // read and configure participation filter
        xp len_mul;
        len_mul = args_recv[3];
        xp len_mul_load[n=1];
        len_mul_load.base = len_mul;
        xp length;
        length = args_recv[4] * len_mul_load[0];

        // do operation
        parallel{
            i in [0, length) loop_color[] = index[i];
            i in [0, length) send_color[] = s0[loop_color[]] * s1[i];
        }
    }
}
\end{lstlisting}

\subsection{Message Passing}
There are two ways to send data between \ProcElem s on the \wse:  1) direct routing configured through Paint, and 2) dynamic routing through Message Passing (\mpi). The CS-3 is the first \wse\ generation to support \mpi. Direct routing has to be configured at compile time and remains fixed throughout execution. This configuration is suitable for simple, well known communication patterns like the routing on the control system and to the worker field. \mpi\ is useful for complex communication patterns and for those that are calculated at runtime. The \mpi\ system uses dimension-ordered routing and a single wavelet header to direct communication between any two arbitrary \ProcElem s. A header is constructed that contains the destination \ProcElem\ position, turn direction information, and termination data. The termination data can specify up to a 5 bit length or termination by a control wavelet. Messaging is atomic and blocking. This makes it possible to support arbitrary slicing in \ProcElem\ dimensions for dense tensor arithmetic and also enables the \DomSpecLang\ to extend to unstructured problems with complex data mappings.

\section{Future Work}
We have used the \virtMachine/\DomSpecLang\ approach for structured grid problems on single wafers to good effect. Given the success we have had and the techniques that we now support in Section \ref{enablingTech}, we 
briefly discuss several research directions that we are currently planned: 

\subsection{Unstructured Grids}
The \DomSpecLang\ is sufficiently flexible to support unstructured problems, especially combined with \mpi\,  provided that the problem mapping respects processor-data locality. We have been actively developing various methods to decompose unstructured meshes accordingly. To this end, we have set up cell-centered and face-centered classes and kernels (built on top of our \DomSpecLang) to support these data constructs. These adaptions of the \DomSpecLang\ enables co-located, unstructured grid modeling on the wafer.

\subsection{Multi-Wafer}
In parallel, we are extending the \virtMachine/\DomSpecLang\ approach to support computation across multiple wafers. This includes running multiple ranks of our \virtMachine/\DomSpecLang\ across a \wse\ cluster and developing requisite inter-\wse\ communication technology to enable fast and efficient communication. We will demonstrate this technology through a Conjugate Gradient solver and preconditioners on unstructured grids under ASCR's Computer Science Competitive Portfolios program.

\subsection{Kernel Fusion}
Currently, we leverage the low 50-60 cycle overhead and the rapid vectorized control system to use fine-grained kernel structures for most programs. This approach performs well with operations on sizable tensors but can be further optimized with kernel fusion. By strategically combining kernels, we can reduce setup costs and thus achieve better strong scaling. Kernel fusion also has the potential to reduce code space required to represent programs and improve performance of critical code sections. Kernel fusion can be enabled either in a context manager at the NumPy level or through automated discovery of kernels to fuse within the \IntRepGraph\ within an optimization step  - or via a combination of both approaches.

\subsection{Advanced Runtime Environment}
We plan to extend our compiler to create host programs that make x86 hosts function as large-memory workers with attached storage. That is, we intend to extend the control system off the wafer to attached hosts to primarily use them as large storage devices. The \wse s are fast enough to fully saturate as many as 12 x86 hosts with IO tasks during computation. If additional resources are left outside of typical IO operations, our computing language can be extended to operate synchronized programs on an array of x86 hosts.

In early testing, we have found that we may need to invest in lossless data compression and/or on live analytics rather than relying on traditional post-processing and analysis techniques used with other architectures. Each WSE is capable of producing over 1 TB/s of data during scientific computing. This volume of data presents a challenge for large-scale computing. It often takes far longer to render data coming off the wafer on traditional visualization hardware than it does to generate the information. Innovation is needed to reduce this burden. 

\subsection{Expanding Hardware Support}

The  \virtMachine/\DomSpecLang\ concept is very powerful as it encompasses all the roles, rules, and responsibilities needed to compile for both spatial architectures as well as conventional unified memory devices. While our main focus has been compiling for the WSE from Python, it should be possible to adapt our compiler to support other spatial architectures and conventional hardware. The \virtMachine\ concept is hardware agnostic and we already run our \IntRepGraph\ through a pure NumPy validation system which is executed on CPU's. In this operational environment, the x86 host takes on the roles given to the control system and each worker type. It is possible to target a common language like C/C++ as an output rather than Tungsten and Paint without violating the concepts in the \virtMachine. Similarly, the controller/worker strategy also encompasses the operating principles for directly attached accelerators like GPU's. In this environment, worker kernels become GPU kernels and the controller actions become host directed kernel launch commands. This should be possible because there is nothing inherent to the \virtMachine\ or \oods\ concepts that would preclude their use on standard, unified memory devices. However, optimizations may take on different character.

\subsection{Multiple Virtual Machines}
This \virtMachine\ and \DomSpecLang\ are not the only possible \virtMachine s and \DomSpecLang s that can be conceived and supported on a flexible machine like the \wse. It is possible to develop similar concepts that are tailored to other tasks. Similarly, not every application needs to consume an entire \wse. For instance, several problems in statistical physics and statistical mechanics would benefit from running many \virtMachine s on the same wafer with varying degrees of interaction. Further, one could develop a framework and language to support many different \virtMachine s running at once to accomplish a larger task.

\section{Conclusions}
We have presented the key operational concepts in MACH: 1) A hardware agnostic, conceptional \virtMachine\ which provides roles, responsibilities, and rules of action and is based on a control system which directs several kinds of workers through \RemProc s. 2) A physical mapping for the \wse\ that enacts the \virtMachine. 3) A \DomSpecLang\ that is composed of allowed data structures (\oods), an \IntLang, an \IntRepGraph, and a memory manager. 4) A compiler system that lowers the \IntRepGraph\ to machine specific code in compliance with the physical mapping. 

We plan to continue enhancing  our \DomSpecLang\ to support unstructured grids and multiwafer systems. We also intend to strengthen our graph optimization techniques to include kernel fusion methods. Additionally, we plan to develop a robust runtime environment with compatible x86 host programs that follow the \IntRepGraph\ and respond as workers in the graph. We also plan to expand to multi-\virtMachine\ operations to support a wide variety of scientific computing workloads including materials modeling, statistical physics, and mechanics applications. 

\section*{Acknowledgments}
The authors would like to thank Marianne Walck, Bryan Morreale, Kirk Gerdes, Tammie Borders, Chris Guenther, MarryAnn Clarke, Kelly Rose, John Crane, Mark Smith, Wei Shi, Brad Shawger, Syndi Credle, and Brian Anderson from NETL for their support. Each of you has played a crucial role in the success of this program and we would not be successful without you.

The authors would like to thank Amirali Sharifian, Nicholas Giamblanco, Milad Hakimi, Kylee Santos, and Michael James from the Advanced Technology Team at Cerebras as well as Andrew Feldman, Andy Hock, Leighton Wilson, and Duncan Hoskinson at Cerebras for their support and help. We appreciate the collaboration we have had so far and look forward to continued successes.

We also appreciate Paola Buitrago, Sergiu Sanielevici, Mei-Yu Wang, and Julian Uran form the Neocortex team at the Pittsburgh Supercomputing Center for access to Neocortex and Bridges 2 on a variety of projects.

Finally, we would like to thank Hal Finkel, Kalyan Perumalla, and David Rabson from the Advanced Scientific Computing Research program at the Office of Science for their support in transitioning our work to unstructured grids and multiple wafers.

This work was supported in part by the U.S. Department of Energy, Office of Science, Office of Advanced Scientific Computing Research's Computer Science Competitive Portfolios program via work authorization to DOE’s National Energy Technology Laboratory.

This work used Neocortex at PSC through allocation CIS250064 from the Advanced Cyberinfrastructure Coordination Ecosystem: Services \& Support (ACCESS) program, which is supported by U.S. National Science Foundation grants \#2138259, \#2138286, \#2138307, \#2137603, and \#2138296.

\bibliographystyle{unsrt}  
\bibliography{00_bibliography}  

\newpage

\section*{Appendix: List of Acronyms}
 \begin{description}
     \item \textbf{\AbsSynTree} -- Abstract Syntax Tree
     \item \textbf{\contextMan} -- Context Manager
     \item \textbf{\DomSpecLang} -- Domain Specific Language
     \item \textbf{\ExecPE} -- Executive Processing Element
     \item \textbf{FIFO} -- First-In, First-Out
     \item \textbf{\GArr} -- Global Array
     \item \textbf{GPU} -- Graphical Processing Unit
     \item \textbf{\GSclr} -- Global Scalar
     \item \textbf{HPC} -- High Performance Computing
     \item \textbf{\IntLang} -- Intermediate Language
     \item \textbf{\IntRepGraph} -- Intermediate Graph Representation
     \item \textbf{\graphInterp} -- Intermediate Graph Representation Interpreter
    \item \textbf{IR} -- Intermediate Representation
     \item \textbf{\LArr} -- Local Array
     \item \textbf{\LSclr} -- Local Scalar
     \item \textbf{\MergePE} -- Merging Processing Element
     \item \textbf{\mpi} -- Message Passing
     \item \textbf{\NetOnChp} -- Network-on-Chip
     \item \textbf{\oods} -- Object Oriented Data Structure
     \item \textbf{\ProcElem} -- Processing Element
     \item \textbf{\RespPE} -- Response Processing Element
     \item \textbf{\RemProc} -- Remote Procedure Call
     \item \textbf{\ULSclr} -- Uniform Local Scalar
     \item \textbf{\virtMachine} -- Virtual Machine
     \item \textbf{\wse} -- Wafer Scale Engine
\end{description}

\end{document}